%% file: main.tex
\documentclass{IEEEoj}
\def\OJlogo{\vspace{-4pt}\hskip-4pt\includegraphics[height=18pt]{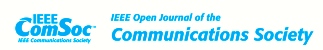}}
\usepackage{cite}
\usepackage{amsmath,amssymb,amsfonts}
\usepackage{algorithmic}
\usepackage{graphicx}
\usepackage{textcomp}
\usepackage[dvipsnames]{xcolor}
\usepackage{url}
\usepackage{todonotes}
\usepackage{float}
\usepackage{subfigure}
\usepackage{siunitx}
\DeclareSIUnit{\gbps}{Gb/s}
\DeclareSIUnit{\mbps}{Mb/s}
\DeclareSIUnit{\gb}{Gb}
\usepackage{tabularx}
\usepackage{csquotes}
\usepackage{booktabs}
\usepackage{flushend}
\usepackage{lettrine}

\usepackage{scalerel}
\usepackage{tikz}
\usetikzlibrary{svg.path}
\definecolor{orcidlogocol}{HTML}{A6CE39}
\tikzset{
  orcidlogo/.pic={
    \fill[orcidlogocol] svg{M256,128c0,70.7-57.3,128-128,128C57.3,256,0,198.7,0,128C0,57.3,57.3,0,128,0C198.7,0,256,57.3,256,128z};
    \fill[white] svg{M86.3,186.2H70.9V79.1h15.4v48.4V186.2z}
                 svg{M108.9,79.1h41.6c39.6,0,57,28.3,57,53.6c0,27.5-21.5,53.6-56.8,53.6h-41.8V79.1z M124.3,172.4h24.5c34.9,0,42.9-26.5,42.9-39.7c0-21.5-13.7-39.7-43.7-39.7h-23.7V172.4z}
                 svg{M88.7,56.8c0,5.5-4.5,10.1-10.1,10.1c-5.6,0-10.1-4.6-10.1-10.1c0-5.6,4.5-10.1,10.1-10.1C84.2,46.7,88.7,51.3,88.7,56.8z};
  }
}
\newcommand\orcidicon[1]{\href{https://orcid.org/#1}{%
\mbox{\begin{tikzpicture}[yscale=-1,transform shape, scale=0.03] % Change scale here
\pic{orcidlogo};
\end{tikzpicture}}}}

\usepackage{hyperref}
\hypersetup{
	colorlinks = true,
	allcolors=blue,
	urlcolor=black,
	linkcolor=black
}
\usepackage[nohyperlinks]{acronym}

\newcommand{\yes}{\checkmark}
\newcommand{\no}{--}

\renewcommand{\thesection}{\Roman{section}}            % Roman for sections (II)
\usepackage{makecell,array,ragged2e}
\newcolumntype{Y}{>{\RaggedRight\arraybackslash}X}   % flexible ragged-right
\newcolumntype{C}{>{\centering\arraybackslash}p{1.1cm}} % compact centered
\newcolumntype{P}{>{\RaggedRight\arraybackslash}p{2.6cm}} % compact paragraph

\newcommand*\ballnumber[1]{\tikz[baseline=(char.base)]{
		\node[shape=circle,fill,inner sep=.5pt] (char) {\textcolor{white}{#1}};}}
\newcommand*\whiteballnumber[1]{\tikz[baseline=(char.base)]{
	\node[shape=circle,fill=white,draw=black,inner sep=.5pt] (char) {\textcolor{black}{#1}};}}

\def\BibTeX{{\rm B\kern-.05em{\sc i\kern-.025em b}\kern-.08em
    T\kern-.1667em\lower.7ex\hbox{E}\kern-.125emX}}

\input{macros}

\begin{document}

\receiveddate{16 March 2026}
\reviseddate{02 April 2026}
\accepteddate{05 April 2026}
\publisheddate{XX April 2026}
\currentdate{08 April, 2026}
\doiinfo{OJCOMS.2026.3682460}

%\jvol{6}
%\pubyear{2025}

\title{P4-TAS: P4-Based Time-Aware Shaper for Time-Sensitive Networking}
%{\footnotesize \textsuperscript{*}Note: Sub-titles are not captured in Xplore and
%should not be used}
%\thanks{We acknowledge support by Open Access Publishing Fund of University of Tübingen.}

\author{FABIAN IHLE$^{\orcidicon{0009-0005-3917-2402}}$, MORITZ FLÜCHTER$^{\orcidicon{0009-0006-6047-5827}}$, AND MICHAEL MENTH$^{\orcidicon{0000-0002-3216-1015}}$\IEEEmembership{(Senior Member, IEEE)}}
\affil{Chair~of~Communication~Networks, University~of~Tuebingen, 72076~Tuebingen, Germany}
\corresp{CORRESPONDING AUTHOR: M. Menth (e-mail: menth@uni-tuebingen.de).}
\authornote{This work was supported in part by the Deutsche Forschungsgemeinschaft (DFG, German Research Foundation) under Grant 544468983, and in part by the Open Access Publishing Fund of the University of Tübingen.}
\markboth{P4-TAS: P4-Based Time-Aware Shaper for Time-Sensitive Networking}{IHLE et al.}

%\author{\IEEEauthorblockN{
%        Fabian~Ihle
%		and Michael~Menth}
%		
%	\IEEEauthorblockA{
%		University~of~Tuebingen,
%		Chair~of~Communication~Networks,
%		72076~Tuebingen,
%		Germany\\
%	}
%	\IEEEauthorblockA{
%        Email: 
%		\{%
%        fabian.ihle,~%
%		michael.menth\}@uni-tuebingen.de
%    }
%}

%\author{\IEEEauthorblockN{1\textsuperscript{st} Fabian Ihle}
%\IEEEauthorblockA{\textit{Chair of Communication Networks} \\
%\textit{University of Tübingen}\\
%72706 Tuebingen , Germany\\
%fabian.ihle@uni-tuebingen.de}
%\and
%\IEEEauthorblockN{2\textsuperscript{nd} Michael Menth}
%\IEEEauthorblockA{\textit{Chair of Communication Networks} \\
%	\textit{University of Tübingen}\\
%	72706 Tuebingen , Germany\\
%	menth@uni-tuebingen.de}
%}

\input{chapters/00-abstract}


\maketitle

\input{chapters/01-introduction}
\input{chapters/02-background}
\input{chapters/03-related_work}
\input{chapters/05-implementation}
\input{chapters/06-evaluation}
\input{chapters/07-conclusion}

\input{glossar}

\bibliography{conferences, literature} 
\bibliographystyle{ieeetr}

\begin{IEEEbiography}
	[{\includegraphics[width=1in,height=1.25in,clip,keepaspectratio] 
		{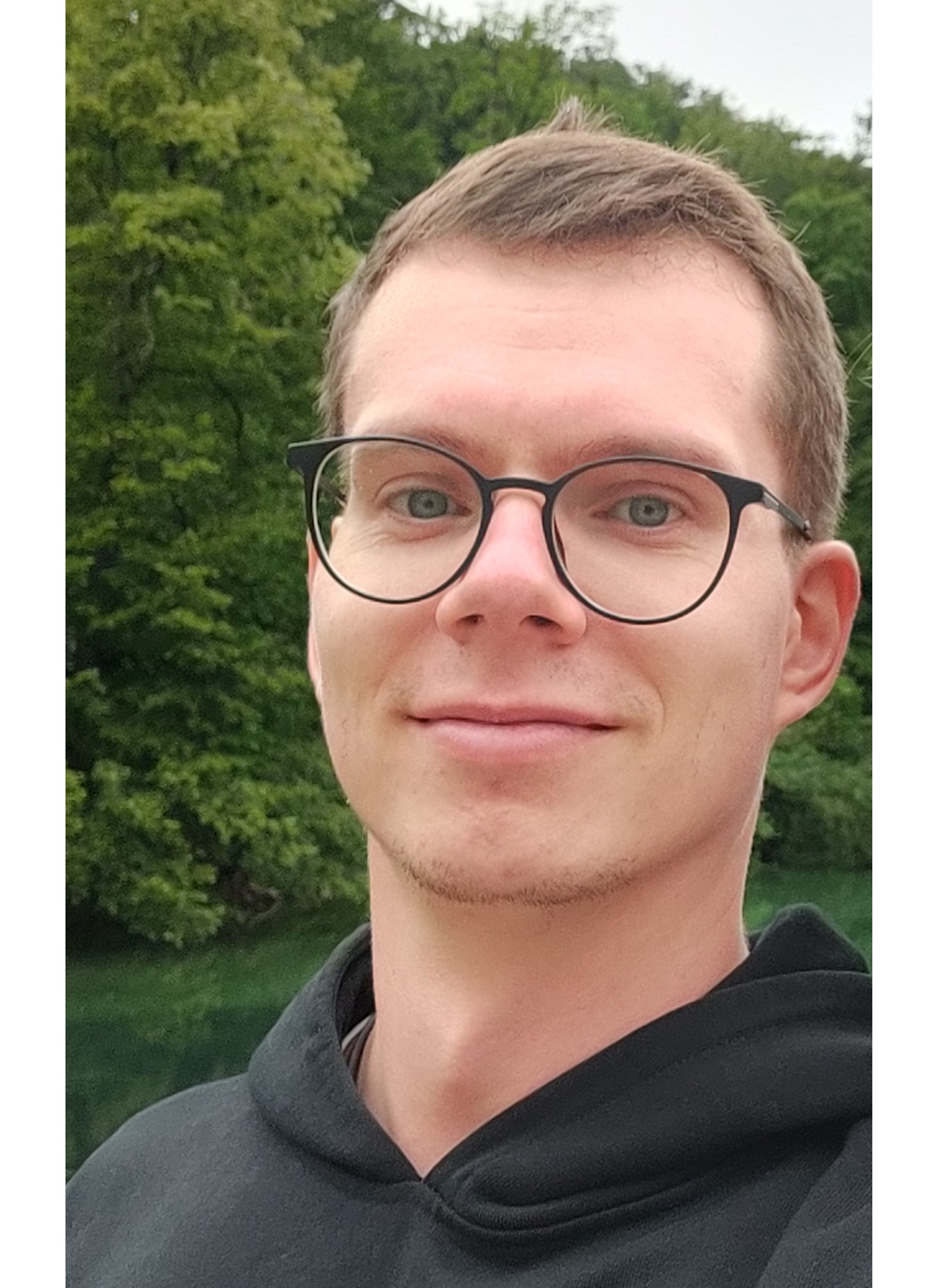}}]{Fabian Ihle }
    received his bachelor's (2021) and master's degrees (2023) in computer science at the University of Tuebingen. Afterwards, he joined the communication networks research group of Prof. Dr. habil. Michael Menth as a Ph.D. student.
    His research interests include software-defined networking, P4-based data plane programming, resilience, and Time-Sensitive Networking (TSN).
\end{IEEEbiography}

\begin{IEEEbiography}
	[{\includegraphics[width=1in,height=1.25in,clip,keepaspectratio] 
		{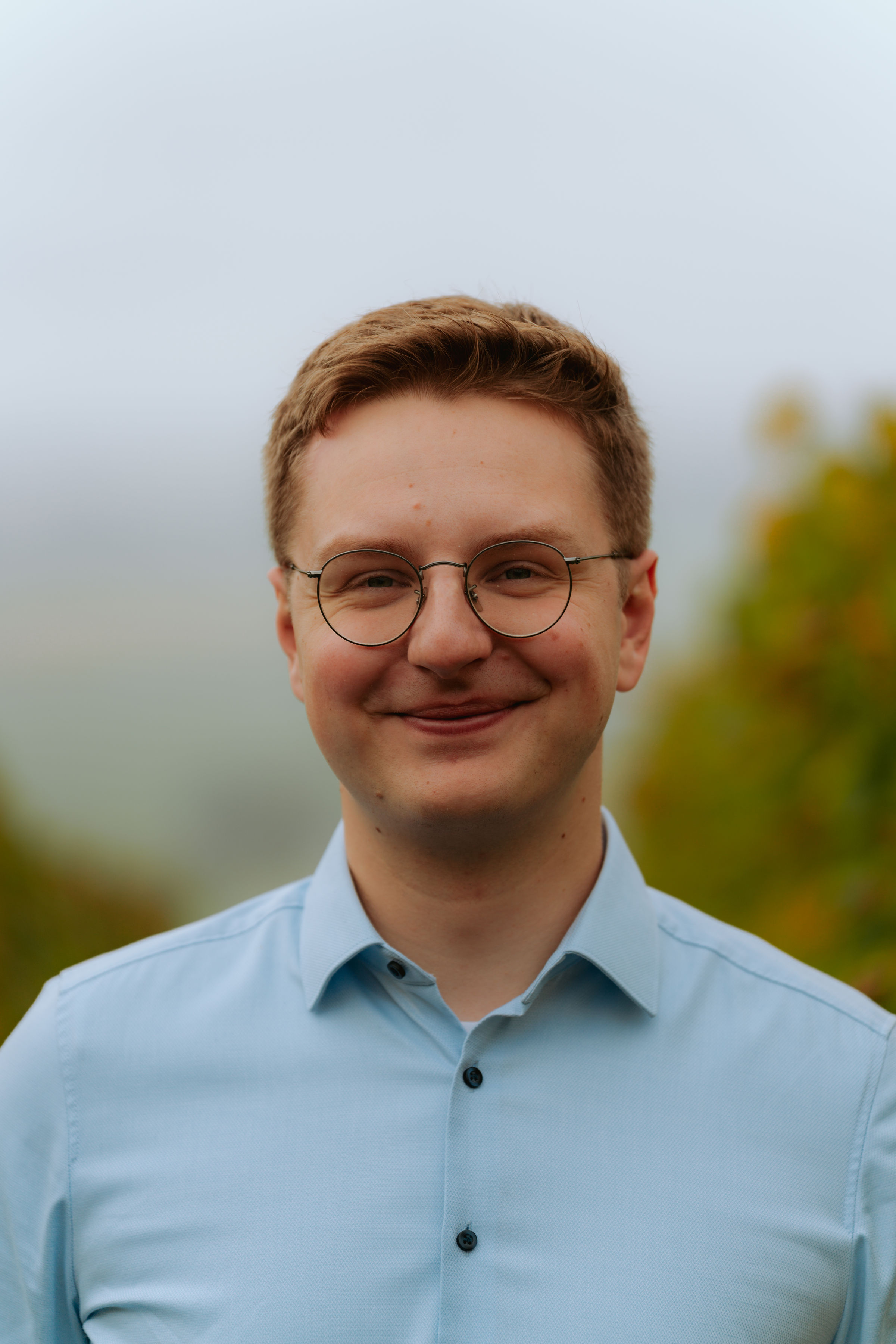}}]{Moritz Flüchter }
    received his bachelor's degree from the DHBW Ravensburg in 2020, and his master's degree in 2023 at the University of Tuebingen.
    He joined the communication networks research group of Prof. Dr. habil. Michael Menth in 2023 as a Ph.D. student.
    His main research interests are data plane programming with P4 and eBPF, routing in low earth orbit (LEO) satellite constellations, and Time-Sensitive Networking (TSN).
\end{IEEEbiography}

\begin{IEEEbiography}
	[{\includegraphics[width=1in,height=1.25in,clip,keepaspectratio]
		{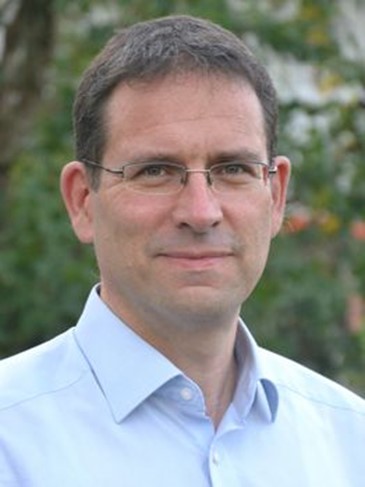}}]{Michael Menth, } (Senior Member, IEEE) is professor at the Department
of Computer Science at the University of Tuebingen/Germany and
chairholder of Communication Networks since 2010. He studied,
worked, and obtained diploma (1998), PhD (2004), and habilitation
(2010) degrees at the universities of Austin/Texas, Ulm/Germany,
and Wuerzburg/Germany. His special interests are performance
analysis and optimization of communication networks, resilience and
routing issues, as well as resource and congestion management. His
recent research focus is on network softwarization, in particular
P4-based data plane programming, Time-Sensitive Networking (TSN),
Internet of Things, and Internet protocols. Dr. Menth contributes
to standardization bodies, notably to the IETF.
\end{IEEEbiography}
 
\end{document}

%% file: macros.tex
\newcommand\citeN\cite

\newcommand\fig[1]{Figure~\ref{fig:#1}}

\newcommand\sect[1]{Section~\ref{sec:#1}}
\newcommand\sects[2]{Section~\ref{sec:#1}--~\ref{sec:#2}}
\newcommand\equa[1]{Equation~\ref{eq:#1}}
\newcommand\tabl[1]{Table~\ref{tab:#1}}

\newcommand\twoeqns[2]{Equations (\ref{eqn:#1}) and~(\ref{eqn:#2})}

\newcommand{\figeps}[3][]{%
 \begin{figure}[htb!]
  \begin{center}
   \leavevmode
      \parbox[t]{#1}{%
        \resizebox{#1}{!}{\includegraphics{figures/#2}}
      }
   \caption{#3\vspace{-0.2cm}}
   \label{fig:#2}
  \end{center}
 \end{figure}
}

\newboolean{makevspace}
\newcommand{\cvspace}[1]{%
    \ifthenelse
        {\boolean{makevspace}}
        {\vspace{#1}}
        {}%
    }

%% file: chapters/00-abstract.tex
\begin{abstract}
Time-sensitive networking (TSN) is a set of IEEE standards that extends Ethernet with real-time capabilities.
Among its mechanisms, the time-aware shaper (TAS) periodically opens and closes egress queues to protect scheduled traffic from lower-priority flows, ensuring low latency and bounded delay.
Deterministic networking (DetNet), standardized by the IETF, provides similar guarantees at Layer~3 and can leverage TSN mechanisms such as the TAS.
Commercially available TSN-capable switches implement TAS in hardware but rarely disclose internal delays in the TAS mechanism itself.
Such delays directly affect scheduling precision, yet information about them is largely unavailable to system designers.
In this work, we present P4-TAS, a P4-based implementation of the TAS on the Intel Tofino\texttrademark~2 switching ASIC that additionally supports per-stream filtering and policing (PSFP) and PTP time synchronization.
First, we design a novel mechanism for periodic queue control that uses a continuous stream of internally generated control frames for time-triggered queue state updates.
To the best of our knowledge, this enables TAS on a P4-programmable ASIC for the first time.
P4-TAS additionally provides an MPLS/TSN translation layer that enables TSN time-based shaping to be applied at the boundary between TSN and DetNet domains, supporting line rates up to 400 Gb/s per port.
Second, we identify and quantify three sources of internal delay that affect the precision of TAS gate transitions, providing transparency that enables more accurate TAS configuration. 
Our evaluation demonstrates a worst-case accumulated internal delay of 86 ns between time slices, which is well below values reported for commercial switches.
Third, we propose a measurement methodology to externally measure TAS time slice accuracy, and introduce gate switching intervals (GSIs) to mitigate overlap between consecutive time slices.
\end{abstract}

%% file: chapters/01-introduction.tex
\section{Introduction}
\label{sec:introduction}
\IEEEPARstart{T}{ime-critical} applications in industrial automation and automotive systems rely on networks that provide deterministic guarantees such as low latency, minimal jitter, and virtually zero packet loss.
To meet these stringent requirements, two complementary technologies have emerged: \ac{TSN} and \ac{DetNet}.
\ac{TSN} is a suite of IEEE 802.1 standards that enhances Ethernet to support real-time communication by introducing mechanisms for traffic shaping~\cite{qbv,qav,qci} and reliability~\cite{cb}.
In contrast, \ac{DetNet} is a Layer 3 technology standardized by the IETF that extends these capabilities to routed networks by enabling bounded latency and high reliability across multiple IP hops~\cite{rfc8655}.

Scheduled traffic is a concept in \ac{TSN} where transmission times of talkers and bridges are coordinated to enable precise timing of frames and bounded latency.
This coordination is called scheduling and yields a network-wide schedule that ensures deterministic forwarding.
Computing such schedules requires knowledge of device-internal timing behavior~\cite{StOs24}, which is rarely disclosed by commercial hardware vendors.
The \ac{TAS} is a \ac{TSN} mechanism which leverages a \ac{TDMA}-like approach to protect scheduled traffic from interference, e.g., by best-effort flows, thereby ensuring low latency and bounded delay.
This is achieved through transmission gates which periodically open and close priority queues.
Further, \ac{PSFP} is a \ac{TSN} mechanism that combines rate and time-based policing to drop out-of-schedule frames.

\ac{DetNet} leverages TSN concepts in combination with technologies such as MPLS or IP.
In \ac{DetNet} deployments, \ac{TSN} can be employed as a sublayer to provide deterministic forwarding in subnetworks.
In practice, \ac{TSN} implementations are typically hardware-based and optimized for bandwidths up to \qty{1}{\gbps}.
However, \ac{DetNet} targets broader use cases, including high-throughput applications and data center backbones.
This hierarchical composition allows for scalable designs that combine the high-speed capabilities of \ac{DetNet} with the precise timing mechanisms of \ac{TSN}.
The integration of both technologies enables sophisticated traffic shaping and reliability mechanisms across backbone infrastructure.

The contribution of this paper is manifold.
First, we design a novel periodic queue control mechanism based on continuously generated internal control frames that trigger queue state updates.
This mechanism addresses two key challenges: P4 does not natively support periodic behavior, and the TAS requires coordinated state changes across all eight egress queues whereas the hardware can only update one queue at a time.
It builds upon our prior P4-PSFP~\cite{IhLi24} work which uses the internal packet generator as a clock source for periodicity, but extends it to simultaneously control multiple queue states, i.e., the transmission gates of the TAS.
To the best of our knowledge, P4-TAS is the first TAS implementation on a P4-programmable switching ASIC.
In addition, P4-TAS incorporates IEEE~Std~1588~\ac{PTP} synchronization with ordinary clock and transparent clock roles, and an MPLS/TSN translation layer based on IEEE Std 802.1CBdb~\cite{8021cbdb} for \ac{DetNet} integration at line rates up to \qty{400}{\gbps} per port.
We evaluate \ac{PTP} accuracy, scalability, and compare P4-TAS against commercial TAS-capable platforms.

Second, we identify and quantify three sources of internal delay that affect the precision of executed TAS schedules: internal traffic generator inaccuracy, queue opening delay, and control frame delay.
These delays do not affect user traffic directly, but rather the \ac{TAS} mechanism itself, e.g., the opening of transmission gates.
Some of those delay sources are inherent to \ac{TAS} implementations in general~\cite{EpOs25} but are typically undisclosed in commercial hardware.
This transparency enables more accurate schedule design~\cite{StOs24}.

Third, we propose a measurement methodology to externally measure the \ac{TAS} time slice accuracy.
The methodology uses a dedicated measurement switch that detects time slice boundaries in the shaped output traffic to determine the actual duration of individual time slices.
It is applicable to other TAS implementations, including commercial black-box switches, as it relies solely on observing output traffic.
We apply this methodology to externally measure the aggregate internal delay bounds from our data plane measurements.
The results confirm consistency between internally measured and externally observed deviations.
Further, they reveal boundary overlap at time slice transitions caused by non-instantaneous queue state changes.
This phenomenon is not specific to P4-TAS but inherent to hardware-based TAS implementations, as independently confirmed by measurements on commercial \ac{TSN} switches by Eppler \textit{et al.}~\cite{EpOs25}.
To mitigate this effect, we propose \acp{GSI}, short explicit time slices that close all queues to cleanly isolate consecutive time slices.
The source code is publicly available on GitHub~\cite{git}.

The rest of the paper is structured as follows.
In \sect{background}, we provide background information on \ac{TSN}, \ac{DetNet}, and the P4 programming language.
In \sect{related_work}, we review related work on combining \ac{TSN} and \ac{DetNet} systems, and on simulations and hardware implementations of those technologies.
\sect{implementation} introduces the P4 implementation, including our system architecture, the PTP implementation, and the P4-TAS mechanism.
In \sect{evaluation}, we evaluate the PTP accuracy and internal timing delays of the TAS mechanism, measure the transmission gate accuracy externally, analyze the scalability, and compare P4-TAS to other implementations.
Finally, in \sect{conclusion}, we conclude the paper.

%% file: chapters/02-background.tex
\section{Technical Background}
In this section, we provide technical background on \ac{TSN}, \ac{DetNet}, and the P4 programming language.
\label{sec:background}

\subsection{Time-Sensitive Networking (TSN)}
\label{sec:background_tsn}
We first give a brief overview of \ac{TSN}, explain scheduled traffic, and then summarize the concepts of the \ac{TAS} and \ac{PSFP}.

\subsubsection{Overview}
\ac{TSN} is a suite of IEEE 802.1 standards that augment traditional Ethernet to support deterministic communication with strict \ac{QoS} guarantees.
\ac{TSN} networks are built from interconnected bridges and end stations.
A data flow, referred to as a TSN stream, originates from a talker (sending station) and is directed to one or more listeners (receiving stations).
A \ac{TSN} stream is identified based on its VLAN tag, its Layer 2 destination address, and optionally other header fields~\cite{cb}.
Before a stream is allowed to transmit, it has to undergo admission control~\cite{qcc}.
This process involves the talker advertising its traffic characteristics, e.g., latency requirements, through a stream descriptor.
The network then decides whether to admit the stream by evaluating resource availability and making reservations accordingly.

\subsubsection{Protecting Scheduled Traffic}
In \ac{TSN}, streams can be scheduled, i.e., their sending times at talkers and bridges are coordinated such that frames experience minimal delay at intermediate bridges.
This coordination is computed offline and yields a network-wide schedule.
The calculation of such schedules is outside the scope of this work.
More information on scheduling in \ac{TSN} can be found in a survey by Stüber \textit{et al.}~\cite{StOs23}.
Time synchronization on a sub-microsecond scale is critical for scheduling in \ac{TSN}.
For that purpose, protocols like \ac{PTP} are employed~\cite{ptp}.

Scheduled streams in \ac{TSN} are typically assigned the highest priority and must be protected from lower-priority traffic, e.g., best-effort traffic.
This ensures that scheduled frames reach each intermediate node at their scheduled times.
\ac{TAS} and \ac{PSFP} are mechanisms to protect scheduled traffic using gating mechanisms.
Both are illustrated in \fig{pdfs/tas_psfp} and explained in the following.

\figeps[\columnwidth]{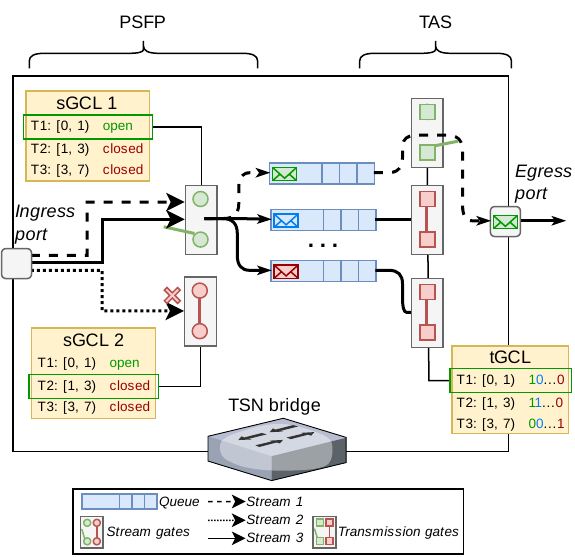}{The \ac{TAS} enables scheduled traffic by gating egress queues while \ac{PSFP} polices stream conformance at the ingress~\cite{qbv,qci}.}

\paragraph{Time-Aware Shaper (TAS)}
\label{sec:background_tas}
The \ac{TAS}, standardized in IEEE Std 802.1Qbv~\cite{qbv}, provides time-based shaping at the egress.
Each egress port provides eight FIFO queues associated with frame priorities from the VLAN tag~\cite{8021q}.
These queues are operated by transmission gates which are controlled by a \ac{GCL}.
A \ac{GCL} is a periodic sequence of entries, each specifying a time slice and a corresponding gate state.
In the \ac{TAS}, we call this the \ac{tGCL}.
Each \ac{tGCL} entry specifies a time slice and an eight-bit vector indicating which of the eight transmission gates are open or closed.
Frames in queues with an open transmission gate are transmitted in FIFO order while frames in queues with a closed transmission gate remain buffered.
After all entries have been processed, the sequence repeats periodically with a cycle length $h$.

\paragraph{Per-Stream Filtering and Policing (PSFP)}
\label{sec:background_psfp}
\ac{PSFP}, standardized in IEEE Std 802.1Qci~\cite{qci}, enforces per-stream conformance at the ingress by combining rate policing with time-based policing.
In this way, \ac{PSFP} ensures adherence to the resource bounds established by admission control.
While rate policing is well established for enforcing bandwidth reservations, time-based policing is required for scheduled traffic to prevent out-of-schedule frames from disrupting the configured timing.
%While rate policing is a well-known mechanism, time-based policing targets scheduled traffic and is the focus of this work.
%For time-based policing, each stream is associated with a stream gate controlled by a periodic \ac{GCL} which we call the \ac{sGCL}.
For time-based policing, streams are associated with stream gates that are controlled by periodic \acp{GCL}, which we call \acp{sGCL}. 
Multiple streams may share the same stream gate and \ac{sGCL}.
The \ac{sGCL} defines the gate state over time and thereby the configured transmission windows of the stream.
Frames arriving outside their scheduled window are dropped immediately, i.e., before queuing, preventing them from consuming reserved resources.

\paragraph{Example and Comparison}
In \fig{pdfs/tas_psfp}, three streams enter the \ac{TSN} bridge which is configured with two stream gates.
Stream 1 and stream 3 are assigned to the first stream gate and stream 2 is assigned to the second stream gate.
Based on the \acp{sGCL}, the first stream gate is currently open while the second is closed.
Accordingly, frames of streams 1 and 3 are queued while frames of stream 2 are dropped by \ac{PSFP}.
Both admitted streams then share the same egress port whose queues are controlled by the \ac{tGCL}.
Frames of streams 1 and 3 are assigned to different egress port queues based on their frame priority.
With the configured \ac{tGCL}, only the transmission gate for queue 1 is currently open, so only frames stored in queue 1 are transmitted.

Stream gates in PSFP differ from transmission gates in TAS in three ways.
First, stream gates apply per stream whereas transmission gates apply per egress port and queue.
Second, an \ac{sGCL} entry defines the state of a single stream gate while a \ac{tGCL} entry defines the states of all eight queues.
Third, closed stream gates drop frames before queueing while closed transmission gates buffer frames in the queue.

\subsection{Deterministic Networking (DetNet)}
The \ac{DetNet} architecture enables real-time applications with extremely low packet loss and a bounded latency~\cite{rfc8655}.
It is standardized by the IETF \ac{DetNet} working group.
DetNet operates at the network layer and provides \ac{QoS} and reliability to lower-layer technologies such as MPLS and TSN.
\acs{DetNet} is applicable to networks under a single administrative control, e.g., to private WANs, or campus-wide networks.

% Data planes
The \ac{DetNet} architecture separates the data plane functions into two sublayers.
First, the service sublayer provides \ac{DetNet} \ac{QoS} mechanisms such as bounded latency and service protection, e.g., by adding sequence number information to packets.
Second, the forwarding sublayer provides connectivity between DetNet processing nodes~\cite{rfc8964}.
Various data plane technologies for \ac{DetNet} exist, e.g., \ac{DetNet} over MPLS~\cite{rfc8964}, and \ac{DetNet} over IP~\cite{rfc8939}.
With \ac{DetNet} over MPLS, the forwarding and service sublayers are identified by MPLS labels, called the \ac{F-Label}, and the \ac{S-Label}.
One or more \acp{F-Label} are used to forward the packet through the \ac{DetNet} domain.
The \ac{S-Label} follows after the \acp{F-Label} and is used to identify the \acs{DetNet} flow.
Based on the identified \acs{DetNet} flow, \ac{QoS} mechanisms are applied.
Further, a \ac{d-CW} follows after the MPLS stack.
This control word contains a sequence number for protection mechanisms of \ac{DetNet}.

To achieve bounded latency, \ac{DetNet} leverages mechanisms defined by the IEEE 802.1 working group such as the \ac{TAS}, combined with bandwidth and buffer reservations at each node.
Standards exist that interconnect \ac{TSN} networks using the \ac{DetNet} MPLS data plane~\cite{rfc9024, rfc9037}.
In \ac{DetNet} MPLS over \ac{TSN}, \ac{DetNet} flows are identified based on the \ac{S-Label} at the \ac{DetNet}/\ac{TSN} domain border and are translated into \ac{TSN} streams.
For that purpose, IEEE Std 802.1CBdb~\cite{8021cbdb} defines an MPLS DetNet flow identification which identifies the \ac{S-Label} and pushes a new VLAN ID.
Then, \ac{TSN} stream identification is applied based on the new VLAN ID.
With those interconnected data planes, \ac{TSN} services such as the \ac{TAS} and \ac{PSFP} can be applied to \ac{DetNet} flows.

\subsection{The Programming Language P4}
\label{sec:background_p4}
\Ac{P4} is a domain-specific programming language to implement custom data planes in P4-programmable switches~\cite{p4}.
A P4 program can manipulate packets and make forwarding decisions to implement custom algorithms.
In the following, we describe the concepts of the P4 pipeline, the packet generator, a feature called \acf{AFC}, and time-based behavior in P4.
A survey by Hauser \textit{et al.} provides more information on P4~\cite{kn}.

\subsubsection{The P4 Pipeline}
P4-programmable switches are called targets and implement a specific architecture.
The Intel Tofino™ 2 switching ASIC is a hardware-based P4 target.
Typically, a P4 architecture follows a pipelined structure.
The pipeline of the \acf{TNA}, the architecture used by the Intel Tofino\texttrademark, is illustrated in \fig{pdfs/pipe_small}.

\figeps[\columnwidth]{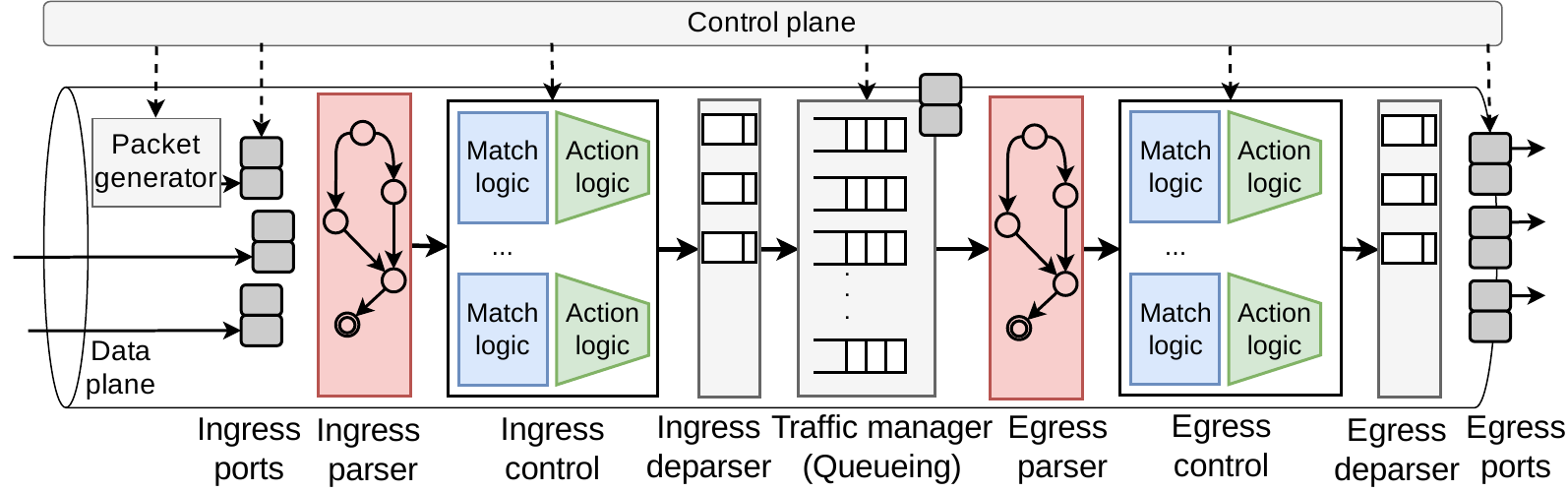}{The pipeline of the \ac{TNA}.}

The \ac{TNA} consists of an ingress block and an egress block, each with a programmable parser, control blocks, and a deparser.
After processing frames in the ingress control block, frames are queued in the traffic manager component of the \ac{TNA}.
This component is configurable but not programmable~\cite{tna}.

Control blocks in a P4 program define the logic of the algorithm.
They leverage metadata for packet processing.
A P4 program defines two different types of metadata.
First, user-defined metadata stores information during the pipeline processing.
Second, intrinsic metadata contains information given by the architecture, e.g., the ingress timestamp of a frame, and the ingress port.
Control blocks are composed of \acp{MAT}.
The concept of a \acs{MAT} is illustrated in \fig{pdfs/mat}~\cite{p4spec} and explained in the following.

\figeps[0.9\columnwidth]{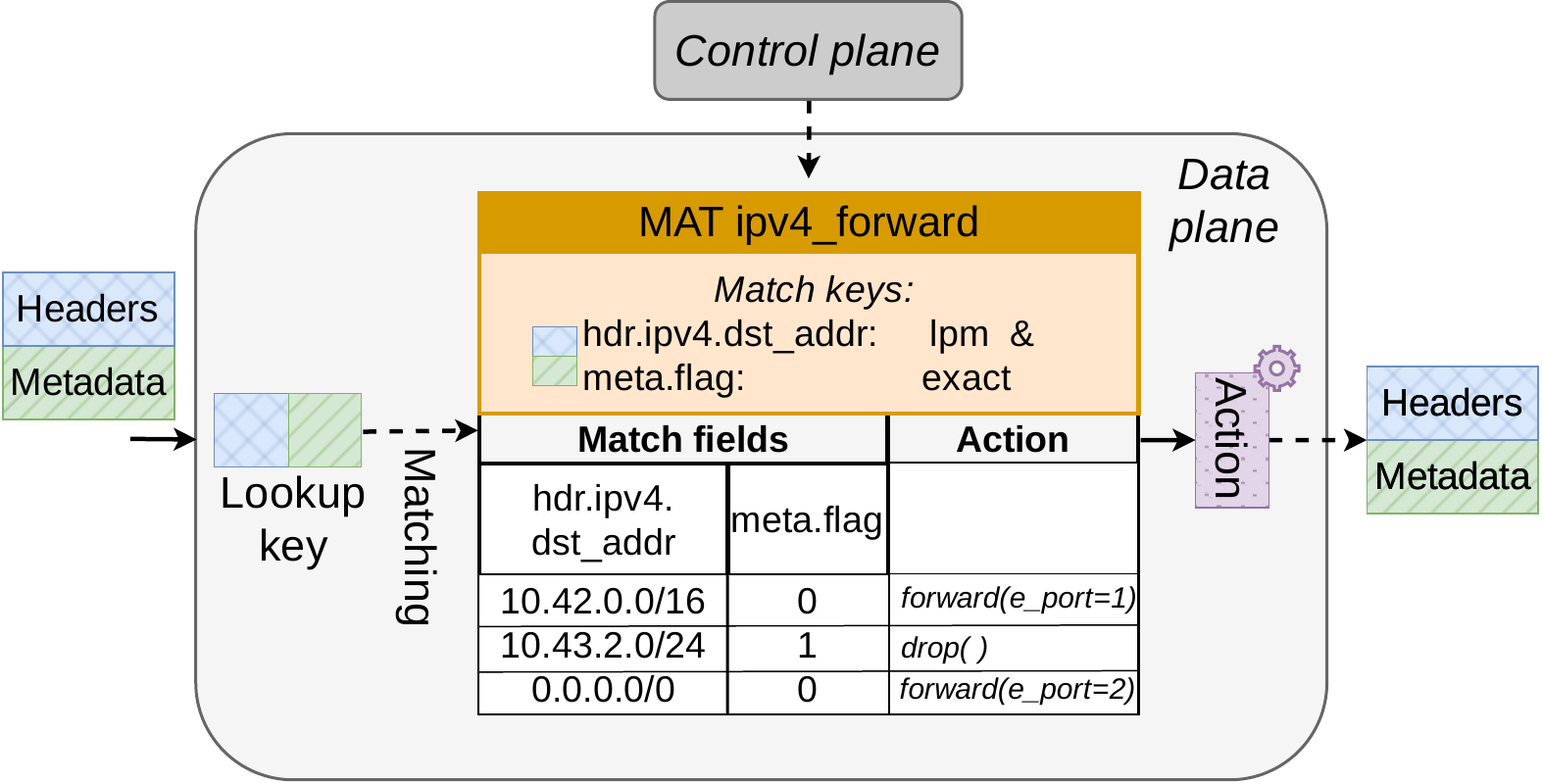}{Concept of \acp{MAT} in P4.}

In a \acs{MAT}, selected packet header fields and metadata form a composite key.
Each packet is matched in the \acs{MAT} according to the selected key fields.
Key fields can have different matching types such as exact, ternary (wildcard bitmask), or range (interval matching).
On a match in the table, an associated action is executed which can manipulate packet data or make a forwarding decision.
The data plane defines the structure of a \ac{MAT}, i.e., the key fields, and the actions.
However, the content of these \acp{MAT} is filled by the control plane.
Further, registers are a commonly used feature in P4 that allow for stateful processing of packets.

P4 control blocks support logical and simple arithmetic expressions but do not support loops to maintain line rate processing.
To enable iterative algorithms, packets can be recirculated.
After a recirculation, modified headers from the first pass are available in the second.
Recirculation introduces delay and requires dedicated ports.
Architectures like the \ac{TNA} offer internal recirculation ports, or can provision physical ports for recirculation.

\subsubsection{The Packet Generator}
The \ac{TNA} provides an internal packet generator which can be configured to generate packets through a dedicated internal port.
Generated packets are processed in the pipeline.
Multiple applications with different triggers, such as a periodic trigger, can be configured to trigger packet generation.
Further, the packet generator can be configured to generate $B$ batches with a batch size of $K$ packets each to enable packet bursts.
A generated packet contains a packet generation header added by the traffic generator.
This packet generation header identifies the application, the batch number, and the packet number in the batch~\cite{tna}.

\subsubsection{Advanced Flow Control (AFC)}
\label{sec:afc}
A feature specific to the Intel Tofino™ 2 is \acf{AFC} which enables control over the queues, i.e., dispatching or holding back frames, of an egress port in the traffic manager.
The queue state is manipulated by writing an \acs{AFC} value into a packet’s intrinsic metadata during pipeline processing.
As this operation must be triggered by an incoming packet, each queue state change is initiated by packet arrival.
A single packet can control exactly one queue.
The \acs{AFC} value is computed based on the egress port, queue ID, and the desired queue state.
Importantly, the controlled queue does not need to correspond to the egress port or queue assigned to the processed packet itself.

\subsubsection{Time-Based Behavior}
\label{sec:background_timing}
The P4 programming language defines data plane logic for describing packet processing operations, e.g., through \acp{MAT}.
However, P4 does not natively provide timing or synchronization capabilities.
These are instead provided by the underlying hardware target.
The Intel Tofino\texttrademark\ supports nanosecond granularity hardware timestamping and clock adjustment APIs for PTP synchronization~\cite{ptp,tna}, as well as hardware features such as the packet generator and \ac{AFC} for queue control.
By combining these platform features with P4's programmable packet processing, time-based execution semantics can be implemented in the data plane.

%% file: chapters/03-related_work.tex
\section{Related Work}
\label{sec:related_work}
In this section, we review related work on the combination of \ac{TSN} and \ac{DetNet} systems, and on simulations and hardware implementations of those technologies. 

\subsection{Combining TSN and DetNet}
The integration of \ac{TSN} and \ac{DetNet} has received considerable attention in recent years due to their critical role in facilitating ultra-low latency communication in 5G networks.
Nasrallah \textit{et al.}~\cite{NaTh19} provide a comprehensive overview of TSN and DetNet technologies, emphasizing their importance for time-critical applications in 5G environments.
Building on this foundation, Abuibaid \textit{et al.}~\cite{AbGh23} conduct a case study that measures the performance of TSN and DetNet in a practical 5G setting.
Furthermore, Wüsteney \textit{et al.}~\cite{BeHe22} propose a latency model for time-sensitive communication traversing networks that integrate TSN and DetNet.
Menendez \textit{et al.}~\cite{MeOl25} present a software-based implementation of the \ac{TAS} using XDP and eBPF.
Further, they integrate \ac{TSN} functionality into \ac{DetNet} environments with an MPLS over UDP/IP data plane.
While their open-source implementation represents a significant step towards \ac{TSN}/\ac{DetNet} integration, their evaluation does not consider internal timing behavior and only measures traffic rates up to \qty{600}{\mbps}.

Despite these advances, challenges remain in realizing efficient hardware implementations that integrate TSN and DetNet functionalities while identifying and quantifying internal timing behavior, which this work aims to address.

\subsection{Simulations and Hardware Implementations of TSN and DetNet}
Numerous simulation frameworks have been developed to model \ac{TSN}~\cite{FaHe19,HeGe16}, \ac{DetNet}~\cite{JiLi19}, or both~\cite{AdIa20}.
In particular, Addanki \textit{et al.}~\cite{AdIa20} offer a simulator that integrates building blocks for \ac{DetNet} at the network layer and \ac{TSN} at the link layer.
Polverini \textit{et al.}~\cite{PoCi25} describe a P4-based \ac{DetNet} implementation for the BMv2 software target leveraging an SRv6 data plane for reliability.
While such simulations are valuable for exploring the interaction between \ac{TSN} and \ac{DetNet}, they do not fully address the challenges of real-world deployment.
Implementing time-sensitive mechanisms in hardware introduces additional complexity due to resource constraints and timing precision requirements.
Ahmed \textit{et al.}~\cite{AhAk24, AkHi24} provide FPGA-based implementations of the \ac{CBS} and \ac{ATS} of \ac{TSN}.
In our prior work~\cite{IhLi24}, we presented a P4-based hardware implementation of the PSFP mechanism on an ASIC.
Several commercial hardware platforms support \ac{TAS} and \ac{PSFP}, such as NXP’s automotive-grade SJA1105TEL switch ASIC~\cite{nxp_1,nxp_2}, and Microchip’s SparX-5i~\cite{sparx_1} and PD-IES008~\cite{microchip_pd1,microchip_pd2} families.
These platforms demonstrate that \ac{TAS} is available in hardware, but published information stops at high-level feature descriptions such as queue counts, \ac{GCL} sizes, or time granularity.
A summary of their capabilities is provided in \sect{eval_compare} and compared against P4-TAS in \tabl{tas_hw_comparison}.

Although these platforms claim nanosecond configuration granularity, reliable queue updates at this scale are not feasible in practice due to undocumented internal delays and hardware limitations.
A recent work by Eppler \textit{et al.}~\cite{EpOs25} identified and quantified such undocumented timing behavior in commercial \ac{TSN} switches.
Their measurements reveal internal scheduling and gate transition delays in the order of hundreds of nanoseconds to several microseconds which is significant for schedule synthesis and can lead to missed transmission windows if not accounted for.
Further, Stüber \textit{et al.}~\cite{StOs24} acknowledge the existence of internal hardware jitter and propose a robust scheduling framework which includes those delays.

In this work, we present a hardware implementation of selected \ac{TSN} and \ac{DetNet} mechanisms on a programmable ASIC.
Unlike prior academic or commercial platforms, our design enables a transparent evaluation of internal delays, thereby offering deeper insights into their behavior and integration in real systems.

%% file: chapters/05-implementation.tex
\section{P4-Based Implementation of TSN Traffic Protection Mechanisms with DetNet Integration}
\label{sec:implementation}
In this section, we describe the implementation of the P4-\ac{TAS} switch incorporating PTP, \acs{PSFP} and the \ac{TAS} mechanism on the Intel Tofino™ 2 switching ASIC.
First, we describe the system architecture and integration into \ac{DetNet} domains.
Then, we present the implementation of the \ac{TAS} mechanism in P4, and the clock synchronization with IEEE Std 1588 PTP.
Finally, we explain improvements to the P4-PSFP implementation, and discuss the portability and generalization of P4-TAS to other platforms.
The source code is publicly available on GitHub~\cite{git}.

\subsection{System Architecture}
The P4-TAS implementation operates as an Ethernet switch that provides TSN functionality.
It performs \ac{TSN} stream identification as defined in IEEE Std 802.1CB~\cite{cb} and applies traffic shaping with the \ac{TAS} as well as policing with \ac{PSFP}.
These mechanisms allow P4-TAS to operate natively inside a \ac{TSN} domain and provide deterministic forwarding for \ac{TSN} streams.
In addition to its role within a pure \ac{TSN} network, P4-TAS can also act as a border element between a \ac{DetNet} domain and a \ac{TSN} domain.
In this case, it processes incoming MPLS-encapsulated \ac{DetNet} flows and translates them into \ac{TSN} streams based on IEEE Std 802.1CBdb~\cite{8021cbdb}.
This enables \ac{DetNet} to leverage the \ac{TSN} sublayer for scheduling and shaping.
The integration scenario is illustrated in \fig{pdfs/detnet}.

\figeps[\columnwidth]{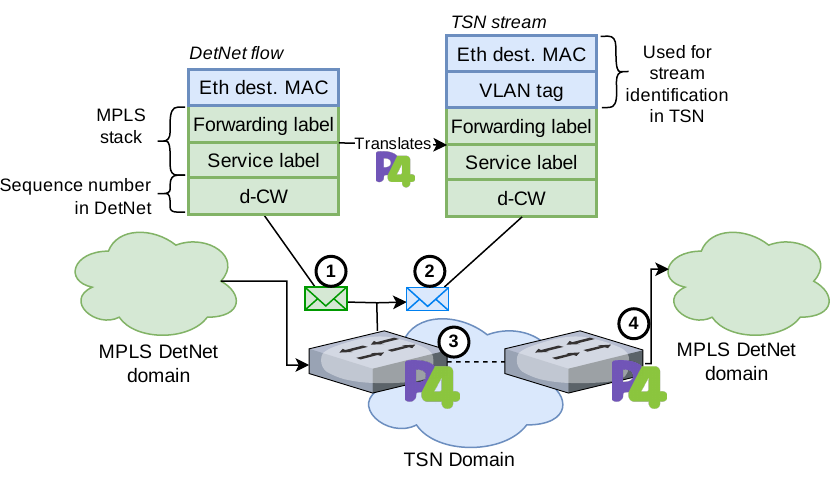}{Placement of P4-TAS in a DetNet environment using a \ac{TSN} sublayer.}

At the ingress to the \ac{TSN} domain (step \whiteballnumber{1}), the P4-TAS switch translates \ac{DetNet} flows into \ac{TSN} streams by pushing a VLAN tag based on the \ac{S-Label}.
Afterwards, \ac{TSN} stream identification is applied using the destination MAC address and the pushed VLAN tag (step \whiteballnumber{2}).
Subsequently, the identified TSN stream is subjected to traffic shaping and policing with TAS and PSFP (step \whiteballnumber{3}), and the frame is forwarded through the TSN domain.
At the egress (step \whiteballnumber{4}), the VLAN tag is removed to restore the original DetNet flow.

\subsection{The Time-Aware Shaper in P4-TAS}
The \ac{TAS} defined in IEEE Std 802.1Qbv periodically opens and closes multiple egress queues according to a \ac{tGCL}.
Periodic behavior, such as in a \ac{GCL}, is not natively supported by P4.
Further, queue states on the Intel Tofino\texttrademark\ 2 can be controlled with \ac{AFC}, but such changes can only be triggered by the arrival of a frame, and each frame can update only a single queue of one egress port.
To implement the \ac{TAS} under these constraints, P4-TAS combines three building blocks: a periodic time model for the \ac{tGCL}, a dedicated stream of continuous \ac{TAS} control frames to trigger \ac{AFC} updates, and a \ac{tGCL} \ac{MAT} that maps control frames to queue state changes.
They are described in the following.
Finally, an overview is given of how they operate together in the pipeline.

\subsubsection{Periodicity of tGCLs}
\label{sec:impl_periodicity}
Modeling the periodicity of \acp{GCL} in programmable P4 hardware is challenging since periodic behavior is not natively supported.
Timestamps in the Intel Tofino\texttrademark\ are absolute, i.e., their values continuously increase, whereas the time slices in \acp{GCL} are relative and follow a periodic pattern.
Thus, each frame’s absolute timestamp must be mapped to its corresponding position within the current \ac{GCL} period. 
While this could be achieved with a modulo operation, such operations are too complex to perform at line rate in the data plane.

In our previous work~\cite{IhLi24}, we described an approach to model the periodicity of \acp{sGCL} in a P4-based \ac{PSFP} implementation.
In this approach, we leveraged the internal packet generator of the Intel Tofino\texttrademark\ as a clock source.
At the end of each \ac{GCL} cycle, the internal packet generator generates a period-completion frame.
The ingress timestamp of this frame is stored in a register and references the timestamp of the last completed period.
For all other frames, i.e., non-period-completion frames, the ingress pipeline subtracts this stored value from the frame's absolute timestamp to obtain a relative timestamp within the ongoing \ac{sGCL} cycle.
In this way, every frame is mapped into a relative time window of one cycle length.
The absolute hardware clock of the switch can be used consistently while the \ac{sGCL} is treated as a repeating list of entries.
We leverage this mechanism to implement the periodicity of \acp{tGCL} for the \acs{TAS}.
However, unlike \acp{sGCL} in \ac{PSFP}, where each \ac{sGCL} entry opens or closes a single stream gate, i.e., admits or drops a frame, a \ac{tGCL} entry must control multiple transmission gates by opening and closing queues.
Thus, the periodicity mechanism serves as the basis for the \acs{TAS}, but additional mechanisms are required to continuously update the gate states of all egress queues during each \ac{tGCL} entry.

\subsubsection{TAS Control Traffic}
\label{sec:impl_tas_traffic}
Queues on the Intel Tofino™ 2 can be opened or closed by processing intrinsic \ac{AFC} metadata of a frame in the pipeline.
In TSN, transmission gates determine whether a queue is eligible for transmission.
Therefore, opening and closing egress queues is functionally equivalent to controlling transmission gates as defined in the \ac{TAS}.
The controlled queue does not need to correspond to the egress port or queue assigned to the frame itself.
A single frame can control exactly one queue of one port of the switch.
In this section, we explain how all queues are controlled in a timely manner to implement the \ac{tGCL}.

To implement \ac{tGCL} queue state changes with \ac{AFC} on the Intel Tofino\texttrademark\ 2, each queue state update must be triggered by the arrival of a frame.
For this purpose, P4-TAS employs the internal packet generator to continuously produce \ac{TAS} control frames.
These frames are generated in back-to-back batches of eight so that each queue is assigned one frame.
Each \ac{TAS} control frame carries intrinsic metadata with the identifiers of the queue and egress port it controls.
Upon arrival, its position in the \ac{tGCL} cycle is calculated based on the frame's arrival timestamp as described in \sect{impl_periodicity}.
Then, the position in the \ac{tGCL} cycle and the intrinsic metadata are matched against the \ac{tGCL} MAT which specifies whether the corresponding queue should be opened or closed at that point in time.

The TAS control frames are minimally sized (\qty{64}{B}) and contain no payload beyond intrinsic metadata.
They are continuously generated with a minimal inter-arrival time to ensure that queue states follow the configured \ac{tGCL} precisely.
This mechanism does not consume bandwidth for user traffic since the internal packet generator and a dedicated internal port are exclusively used for the TAS control traffic.
In practice, there is a short delay between consecutive \ac{TAS} control frames.
Since a queue can only change state when its associated control frame arrives, delayed opening or closing may occur.
We evaluate the impact of this behavior in \sect{eval_batch_delays}.
The \ac{TAS} control traffic uses one of the 16 available periodic packet generator applications.
The remaining 15 applications are available for period-completion frames, allowing up to 15 concurrent \ac{GCL} periods to be configured.

The internal packet generator operates independently of external traffic and does not share resources with user-facing ports.
If the packet generator were to stall, queue states would remain frozen in their last configured state.
Such a condition could be detected through periodic control plane health checks, e.g., by continuously reading and validating the last period-completion timestamp.

\subsubsection{The tGCL MAT}
\label{sec:impl_mat}
The \ac{tGCL} is encoded as a \ac{MAT} in the egress pipeline and is shown in \fig{pdfs/gcls-tgcl_mat}.
\ac{TAS} control frames from \sect{impl_tas_traffic} are matched against it.
\figeps[\columnwidth]{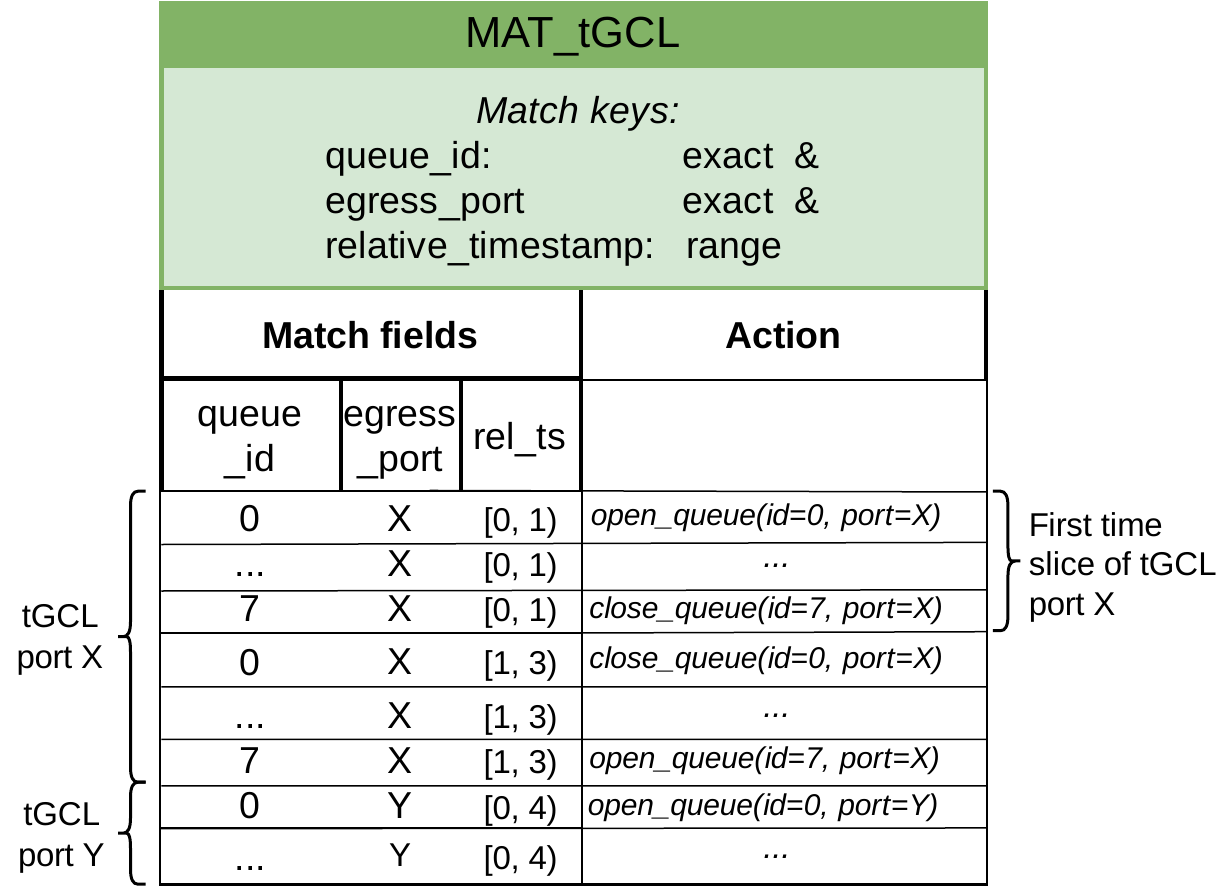}{The \acs{tGCL} of the \acs{TAS} is modeled as a \acs{MAT}. The generated \acs{TAS} control traffic is matched in this table. As an action, the corresponding queue is opened or closed.}

Each entry in the \ac{MAT} corresponds to one of the eight queues of a \ac{tGCL} entry, i.e., eight \ac{MAT} entries per \ac{tGCL} entry are required.
The entry specifies whether a queue should be currently open or closed.
The lookup key is composed of the relative timestamp in the \ac{tGCL} which is calculated according to \sect{impl_periodicity}, the queue identifier, and the egress port.
The \ac{MAT} action writes a precomputed \ac{AFC} value which encodes the queue, the egress port, and the state, into the frame’s intrinsic metadata.
This triggers the queue state update.
The queue state update has a small delay which is evaluated in \sect{eval_queue_delay}.

\subsubsection{Processing Paths in P4-TAS}

The mechanisms in \sects{impl_periodicity}{impl_mat} operate together within the P4-TAS pipeline as illustrated in \fig{pdfs/gcls-pipeline_all}.

\figeps[\columnwidth]{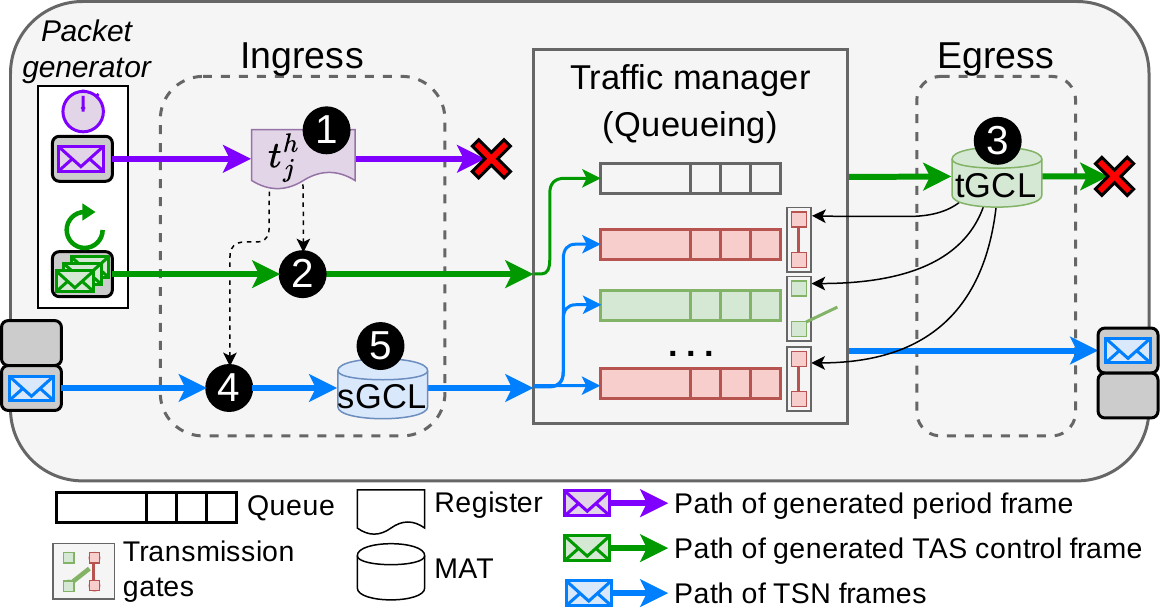}{Overview of the P4-TAS pipeline.}

First, generated period-completion frames mark the boundaries of the \ac{tGCL} and \ac{sGCL} cycles and maintain the reference for relative timestamp calculation in step \ballnumber{1}.
Here, a single frame is generated at the end of each period with duration $h$, and its timestamp of the $j$-th period $t^h_j$ is stored in a register for subsequent processing.
Afterward, those frames are dropped.

Second, \ac{TAS} control frames are continuously generated by the packet generator with a minimal inter-arrival time.
For those frames, the timestamp relative to the last elapsed period of the \ac{tGCL} is calculated in step \ballnumber{2}.
This timestamp is used to match the \ac{TAS} control frame to the corresponding entry of the \ac{tGCL} \ac{MAT}.
After queuing the \ac{TAS} control frame in a dedicated queue of the traffic manager, the \acs{AFC} mechanism is applied in the egress in step \ballnumber{3}.
Here, the corresponding queue is opened or closed based on the current \ac{tGCL} entry using the \acs{MAT} described in \sect{impl_mat}.
Afterward, the frames are dropped.

Third, \ac{TSN} data frames are policed in the ingress pipeline by the \ac{PSFP} mechanism to enforce conformance with admitted rates and transmission times.
Each frame is first matched to a \ac{TSN} stream using the stream identification function defined in IEEE Std 802.1CBdb and then assigned to the corresponding stream gate.
Next, the timestamp relative to the last elapsed period of the \ac{sGCL} is calculated in step \ballnumber{4}, and \ac{PSFP} is applied in step \ballnumber{5}.
In this step, frames are policed and either dropped or forwarded to the appropriate priority queue.
The queues are either open or closed based on the \acs{tGCL}.
Frames are forwarded in a FIFO manner as soon as their queue opens.

\subsection{Clock Synchronization with IEEE Std 1588 PTP}
\label{sec:impl_ptp}
\ac{PTP}, defined in IEEE~Std~1588~\cite{ieee1588}, enables sub-microsecond clock synchronization by exchanging timestamped messages between network nodes and a grandmaster clock.
The Intel Tofino\texttrademark\ 2 ASIC provides hardware support for \ac{PTP}.
This includes hardware timestamping with nanosecond granularity, and data plane API endpoints to retrieve timestamps and adjust the local clock~\cite{tna}.
However, \ac{PTP} message handling in the data and control planes must be implemented by the user.
We implement the two-step \ac{E2E} delay measurement mechanism defined in IEEE~Std~1588.
The \ac{P2P} mechanism defined in IEEE~Std~802.1AS~\cite{as} (gPTP) can increase the \ac{PTP} synchronization accuracy over multiple hops~\cite{DuYa24}, but increases the implementation complexity.
A typical deployment using the \ac{E2E} mechanism is shown in \fig{pdfs/ptp}.

\figeps[\columnwidth]{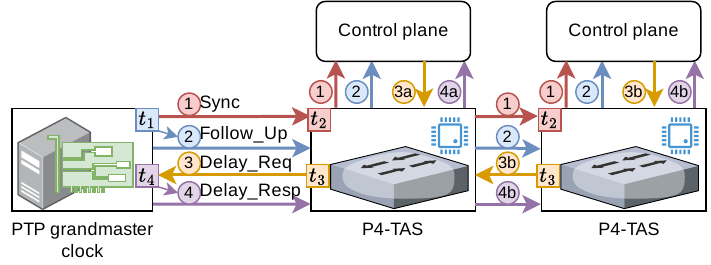}{Deployment of IEEE~Std~1588 \ac{E2E} \ac{PTP} synchronization with P4-TAS acting as ordinary and transparent clock. The grandmaster synchronizes the first switch, which replicates \ac{PTP} messages with residence time correction to a downstream switch.}

The P4-TAS switch is connected to a \ac{PTP} grandmaster clock, e.g., an external server running \texttt{ptp4l} and a NIC with hardware timestamping, which periodically sends PTP messages to the switch.
For \ac{PTP} traffic, isolated physical ports and queues are used so that synchronization is independent of other traffic.
The implementation supports two \ac{PTP} roles: an \acf{OC} and a \acf{TC}.
Both roles can be active simultaneously on the same switch.
They are described in the following.

\subsubsection{Ordinary Clock (OC)}
\label{sec:impl_ptp_oc}
In the \ac{OC} role, the local clock of the Tofino\texttrademark\ ASIC is synchronized to the \ac{PTP} grandmaster using the IEEE~Std~1588 \ac{E2E} delay measurement mechanism.
The message exchange proceeds as follows.
First, the grandmaster sends a \texttt{Sync} message.
The P4-TAS data plane captures its ingress hardware timestamp $t_2$ at the physical interface and appends it to the packet.
This packet is then forwarded to the control plane via an internal \qty{10}{\gbps} interface.
The grandmaster sends a \texttt{Follow\_Up} message which carries the egress timestamp~$t_1$ of the \texttt{Sync} message at the grandmaster.
This message is also forwarded to the control plane.
After receiving~$t_1$ and~$t_2$, the control plane generates a \texttt{Delay\_Req} message which is forwarded through the data plane to the grandmaster.
The egress hardware timestamp~$t_3$ of this message at the physical interface of the switch is captured through the Tofino\texttrademark\ data plane API and recorded by the control plane.
The grandmaster responds with a \texttt{Delay\_Resp} message containing the ingress timestamp~$t_4$ of the \texttt{Delay\_Req} at the grandmaster.

With all four timestamps available, the control plane computes the path delay~$d$ and the offset from the master~$o$ according to IEEE~Std~1588:
\begin{align}
    d &= \frac{(t_2 - t_1) + (t_4 - t_3)}{2},
    \label{eqn:path_delay} \\
    o &= (t_2 - t_1) - d.
    \label{eqn:offset}
\end{align}

Offsets are calculated continuously for every \ac{PTP} message exchange.
The first computed offset typically reflects a large time domain mismatch between the local ASIC clock and the \ac{PTP} time.
This initial offset is stored as a baseline and subtracted from all subsequent measurements to isolate the actual synchronization drift.
The clock is then adjusted using two complementary mechanisms provided by the Tofino\texttrademark\ data plane API: a global timestamp offset for time correction, and a clock increment adjustment to compensate for frequency drift.

\subsubsection{Transparent Clock (TC)}
The \ac{TC} role enables PTP synchronization of downstream devices by forwarding PTP messages through the P4-TAS switch with residence time correction.
P4-TAS implements an \ac{E2E} \ac{TC} as defined in IEEE~Std~1588.
For that purpose, \texttt{Sync} and \texttt{Follow\_Up} messages from the grandmaster are replicated and sent to downstream ports.
For each PTP packet forwarded to a downstream receiver, the data plane computes the switch residence time as the difference between the egress and ingress hardware timestamp.
This value is added to the correction field of forwarded \ac{PTP} messages to account for the processing delay in the switch.
A downstream \ac{OC} subtracts the accumulated correction fields from the measured delays in \twoeqns{path_delay}{offset} to compensate for residence times in upstream switches.

\subsubsection{GCL Activation}
\label{sec:gcl_activation}
For synchronized TAS operation, GCL cycles on different switches must start at the same global time.
In P4-TAS, this is achieved by delaying the start of the GCL cycle until the local PTP-synchronized clock reaches a preconfigured absolute start time.
Before activation, all GCL entries and configurations are installed in the data plane. 
The control plane is then configured with the desired start time, e.g., the current grandmaster time plus a defined offset.
Once the local clock reaches this time, the control plane activates the internal packet generator, which begins the GCL cycle.
In our implementation, the start time is configured manually.
In production deployments, standardized mechanisms such as the centralized network configuration (CNC) model defined in IEEE Std 802.1Qcc, using protocols such as NETCONF/YANG, can automate the distribution of GCL configurations and start times to all bridges in the network.

\subsection{Improvements to P4-PSFP}
P4-TAS incorporates the previous P4-PSFP implementation~\cite{IhLi24}.
The PSFP components stream filter, stream gate, and flow meter are implemented according to IEEE Std 802.1Qci~\cite{qci}.
The functionality of P4-PSFP has been extensively evaluated in~\cite{IhLi24}.
In this section, we describe improvements to P4-PSFP that eliminate recirculation, and increase the time resolution of \acp{GCL}.

\subsubsection{Eliminating Recirculation}
P4-PSFP recirculates \acs{TSN} traffic for two reasons.
First, calculating the relative position in a \acs{sGCL} does not fit in a single pipeline iteration.
Second, the optional maximum frame size filter defined in IEEE Std 802.1Qci~\cite{qci} requires frame size info only available in the egress block while drops must occur in the ingress block.
Thus, recirculation is necessary in P4-PSFP, adding a known constant delay.
For P4-TAS, we ported the implementation of P4-PSFP from Intel Tofino™ to Tofino™ 2 where the larger pipeline allows the \acs{GCL} position to be computed in one pass.
We also removed the optional maximum frame size filter, eliminating the need for recirculation.
If required, the filter can be re-added, at the cost of a recirculation.

\subsubsection{Increased Time Resolution Using Range-to-Ternary Conversion}
\label{sec:range-to-ternary}
\ac{sGCL} entries in P4-PSFP are modeled as \ac{MAT} entries with the range matching type.
However, the range matching type is limited in the \ac{TNA} such that at most \qty{20}{bits} can be matched.
Timestamps in the \ac{TNA} are \qty{48}{bits} with nanosecond granularity.
Therefore, in P4-PSFP, \qty{20}{bits} are cut out of the middle of the timestamp to enable the range matching type and enable an appropriate time resolution.
Thus, \acp{GCL} have a minimum resolution of \qty{2}{\us} and a maximum resolution of approximately \qty{4}{\second}.
\ac{GCL} entries with a lower resolution, or \acp{GCL} that last longer cannot be defined in P4-PSFP.
However, due to hardware limitations, P4-TAS requires small intervals between \ac{tGCL} entries where a minimum resolution of \qty{2}{\us} is too large.
This is further elaborated in \sect{guard_bands}.
Therefore, we employ an algorithm called range-to-ternary conversion~\cite{GuMc01} to increase the resolution of time slices.
This algorithm allows modeling a single range entry using multiple ternary entries.

The algorithm takes an integer range $[L,R]$ representing a time slice and breaks it down into the smallest possible set of prefixes that collectively cover the entire range.
It does this by repeatedly selecting the largest prefix starting at the current lower bound that remains fully within the range.
These selected blocks together ensure complete coverage of the interval~\cite{GuMc01}.
Some example conversions are given in \fig{pdfs/range_to_ternary_example}. 
Each block in \fig{pdfs/range_to_ternary_example} denotes a ternary entry that covers parts of the range.
The $*$ denotes a “don’t care” bit, meaning the bit can take either value 0 or 1.

\figeps[\columnwidth]{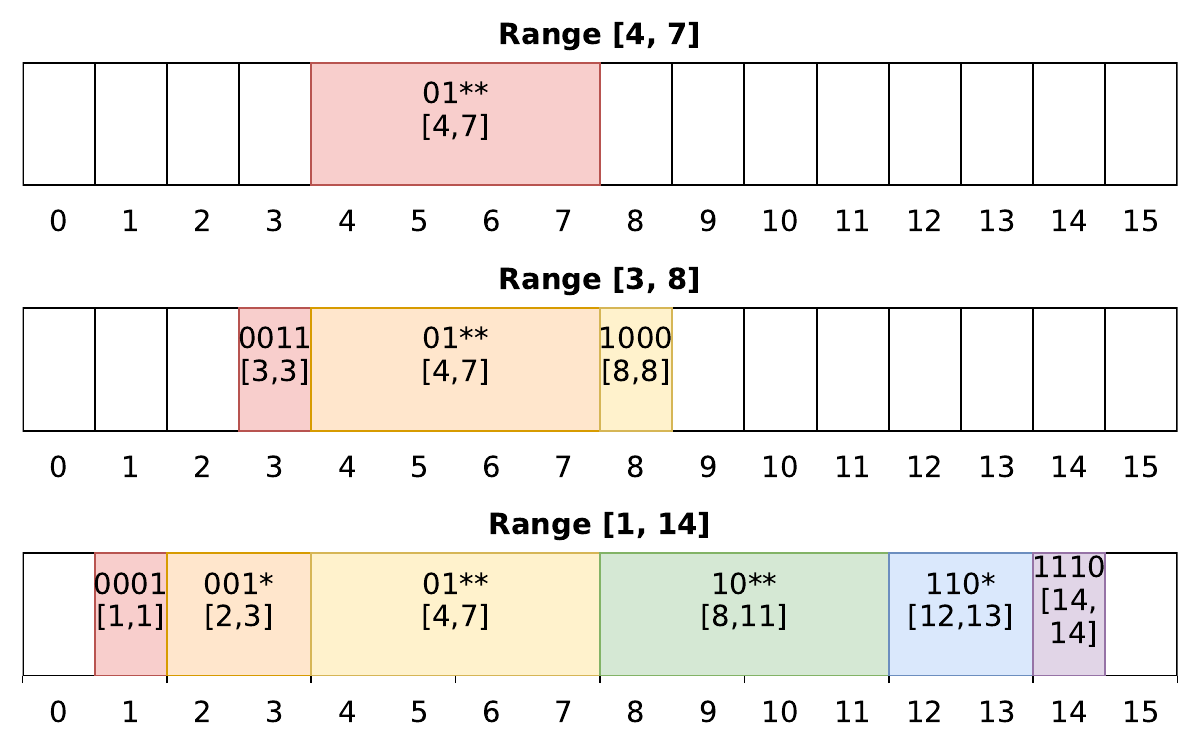}{Example of range-to-ternary mappings for three different ranges~\cite{GuMc01}.}

In a \ac{GCL}, time slices are defined as consecutive, non-overlapping ranges.
Under these constraints, Sun~\cite{Su11} has proven that the solution is both correct and unique.

With this algorithm, \acp{GCL} have a resolution of \qty{1}{\nano\second} to \qty{78}{\hour}.
The upper bound of \qty{78}{\hour} exceeds the requirements of \ac{GCL} periods by far and is not necessary in \ac{TSN} networks.
However, the full \qty{48}{bits} timestamp range is available for matching, and reducing the resolution does not have a benefit.
The number of ternary table entries required by this conversion algorithm to model \acp{GCL} is evaluated in \sect{scalability}.

\subsection{Portability and Generalization}
\label{sec:portability}
While P4-TAS is implemented on the Intel Tofino\texttrademark\ 2, the core mechanisms can be adapted to other platforms. We identify which components are Tofino-specific and which concepts generalize.

Our implementation relies on two platform-specific capabilities.
First, the internal packet generator provides a programmable source of periodic events with nanosecond-level timing. 
Second, \ac{AFC} enables dynamic queue state manipulation by writing intrinsic metadata during packet processing.
These features are leveraged in our implementation but are not fundamental requirements.

The core requirement is periodic behavior in the data plane.
While we use the packet generator as a clock source, alternative platforms could achieve this through hardware timers or external triggers.
The key principle of using timed events to mark \ac{GCL} boundaries and trigger queue updates remains applicable.
Queue control via packet-triggered events (\ac{AFC} in our case) could be implemented through vendor-specific APIs, or traffic manager interfaces on other platforms.
Hardware timestamping with nanosecond granularity is essential for \ac{TSN} but widely available in modern switching hardware.
\ac{PTP} support requires means to extract timestamps and adjust the local clock.

Beyond timing primitives, the platform must support programmable packet processing to implement custom data plane logic such as \ac{GCL} matching, relative timestamp computation, and stream identification.
While P4 provides this through \acp{MAT}, equivalent functionality is available in other programmable hardware architectures such as FPGA-based designs.

In summary, while P4-TAS leverages Tofino-specific features, the core design patterns of event-driven queue control and timestamp-based \ac{GCL} matching generalize to other hardware platforms.

%% file: chapters/06-evaluation.tex
\section{Evaluation of P4-TAS}
\label{sec:evaluation}
In this section, we evaluate the P4-TAS implementation.
First, we assess the accuracy of the \ac{PTP} synchronization over multiple hops.
Then, we identify and quantify internal delays including the traffic generator accuracy, the queue opening delay, and the \acs{TAS} control frame delay.
Next, we externally measure the duration of \ac{tGCL} entries to confirm consistency with the internally measured delays.
For that purpose, we introduce \acp{GSI} to mitigate transitional behavior between \ac{tGCL} entries resulting from the queue opening delay.
Further, we assess the scalability of P4-TAS by analyzing the number of supported \ac{tGCL} and \ac{sGCL} entries, the range-to-ternary conversion overhead, and the maximum number of streams for identification of \ac{DetNet} and \ac{TSN} flows.
Finally, we compare P4-TAS to available TAS implementations.

\subsection{PTP Synchronization Accuracy}
\label{sec:eval_ptp}
\begin{figure}[htb]
  \leavevmode
  \begin{center}
    \subfigure[Switch 1.]{
      \label{fig:ptp_sw1}
      \parbox[t]{0.46\columnwidth}{%
        \resizebox{0.46\columnwidth}{!}{%
          \includegraphics{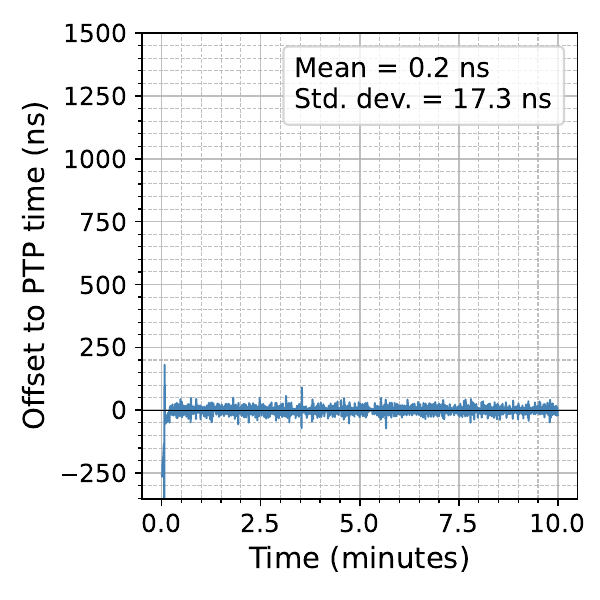}%
        }%
        \vspace{-1cm}
      }
    }
    \subfigure[Switch 2.]{
      \label{fig:ptp_sw2}
      \parbox[t]{0.46\columnwidth}{%
        \resizebox{0.46\columnwidth}{!}{%
          \includegraphics{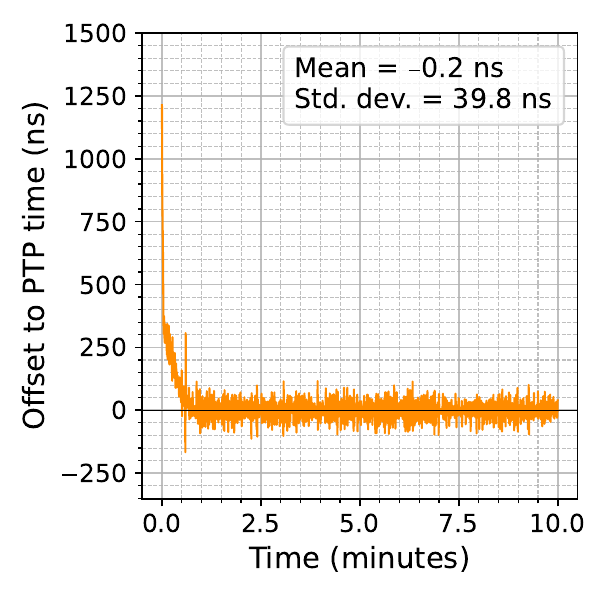}%
        }%
        \vspace{-1cm}
      }
    }
  \end{center}
  \vspace{0cm}
  \caption{Measured offset from \ac{PTP} time for two chained P4-TAS switches: Switch~1 is directly synchronized to the grandmaster, Switch~2 is synchronized via Switch~1 acting as \acs{TC}.}
  \label{fig:eval_ptp}
\end{figure}

To validate the \ac{PTP} implementation described in \sect{impl_ptp}, we measure the offset from the grandmaster clock in two chained P4-TAS switches over \qty{10}{minutes}, i.e., the synchronization accuracy.
The testbed follows the deployment shown in \fig{pdfs/ptp}: a \ac{PTP} grandmaster clock synchronizes the first P4-TAS switch (\ac{OC}), which additionally acts as a \ac{TC} toward a second downstream P4-TAS switch (\ac{OC}).
The grandmaster runs \texttt{ptp4l} on a server equipped with a Mellanox ConnectX-5 NIC with hardware timestamping, using a free-running clock as the time source.
The grandmaster transmits two \texttt{Sync} messages per second.
Each switch computes its offset from the \ac{PTP} grandmaster time as described in \sect{impl_ptp_oc}, and the values are recorded over the duration of the experiment.
The results are shown in \fig{eval_ptp}.

Both switches show a large initial offset which converges to a steady state within five seconds for the first switch, and within a minute for the downstream second switch.
The downstream switch takes longer to converge since it started with a higher offset from the grandmaster time.
After convergence, the first switch achieves a mean offset of \qty{0.2}{\ns} (std.\ dev.\ \qty{17.3}{\ns}), while the second switch achieves \qty{-0.2}{\ns} (std.\ dev.\ \qty{39.8}{\ns}).
Both switches maintain a stable offset throughout the measurement.
The downstream switch shows a higher steady-state standard deviation of \qty{39.8}{\ns} which is typical for multi-hop E2E \ac{PTP} synchronization via a \ac{TC} as identified by Rezabek \textit{et al.}~\cite{ReHe22}.
This offset is comparable to other hardware-assisted PTP implementations~\cite{KySp19}, confirming sub-microsecond synchronization across the \ac{TAS} domain.
Since a dedicated physical port and queues are used for PTP traffic, other traffic has no impact on the synchronization accuracy.
We chose the E2E delay measurement mechanism for its simpler implementation requirements.
The P2P mechanism defined in gPTP can reduce hop-induced offset accumulation~\cite{DuYa24} but adds implementation complexity.
Exploring P2P synchronization is left for future work.

If \ac{PTP} synchronization drifts, \ac{GCL} cycles on neighboring nodes become misaligned, potentially causing frames to arrive outside their scheduled transmission times.
Scheduling approaches that model synchronization drift as bounded intervals, such as those by Stüber \textit{et al.}~\cite{StOs24} and Craciunas \textit{et al.}~\cite{CrOl21}, can compute guard bands to tolerate such deviations.
Our measured steady-state offsets with a standard deviation up to \qty{39.8}{\ns} provide the required input parameters for these approaches.
 
\subsection{Internal Timing Delays of the TAS Mechanism}
\label{sec:eval_internal_delays}
Prior work has identified that internal hardware delays and jitter affect TSN scheduling correctness~\cite{StOs24, EpOs25}, and some vendors partially disclose such values~\cite{sparx_1}.
The specific sources of these delays and their individual contributions are rarely characterized.
Franco \textit{et al.}~\cite{FrZa24} profile the processing latency behavior of the Intel Tofino\texttrademark\ ASIC.
However, beyond processing delays, additional delay sources exist within \ac{TSN} bridges that are not typically disclosed~\cite{EpOs25}.
These delays do not directly delay user traffic, but affect \ac{TAS} control operations such as queue state changes.
In this section, we identify and quantify such internal delay sources in our P4-TAS implementation on the Intel Tofino\texttrademark\ 2 platform.
While the measurement results are specific to this platform, the sources of these delays are also present in other hardware~\cite{EpOs25, sparx_1}.
These measurements characterize local timing behavior which is independent of PTP synchronization.

First, we evaluate the accuracy of the internal traffic generator which affects the timing of period-completion frames.
We then analyze the queue opening delay of the \ac{AFC} mechanism.
Finally, we measure a delay introduced by the packet generator used for \ac{TAS} control frames, and give a summary of the measurements.

\subsubsection{Traffic Generator Accuracy}
\label{sec:eval_traffic_gen}
P4-TAS uses the internal packet generator to signal the completion of each \acs{tGCL} cycle with a configured period $h$ as described in~\sect{impl_periodicity}.
A period-completion frame is generated every $h$~ns, and the timestamp of the $j$-th period, denoted as $t^h_j$, is stored in a register.
Due to limitations of the packet generator, small timing deviations may occur.
To quantify this effect, we measure the difference between the timestamps of consecutive period-completion frames, i.e., $t^h_{j+1}$ and $t^h_j$, relative to the configured period $h$.
The deviation $\hat{\delta}_{\text{TG}}$ is defined in \equa{delta_tg}:
\begin{align}
    \hat{\delta}_{\text{TG}} &= (t^h_{j+1} - t^h_j) - h\label{eq:delta_tg}.
\end{align}

This value is recorded as a time series in a register in the data plane.
Based on use cases identified by Stüber \textit{et al.}~\cite{StEp24}, we select representative periods $h$: \qty{500}{\us} for factory automation, \qty{2}{\ms} for industrial isochronous traffic, and \qty{128}{\ms} for aerospace applications.
Additionally, we include \qty{10}{\us} to test short-period behavior, \qty{400}{\us}, \qty{800}{\us} and \qty{1}{\ms} to cover the \qty{400}{\us} to \qty{2}{\ms} industrial range, and \qty{32}{\ms}, \qty{64}{\ms}, and \qty{96}{\ms} to cover the range up to \qty{128}{\ms}.
For each period, we record the timestamps of 16000 period-completion frames.
\fig{pdfs/boxplot_clock_offset} shows the results.

\figeps[\columnwidth]{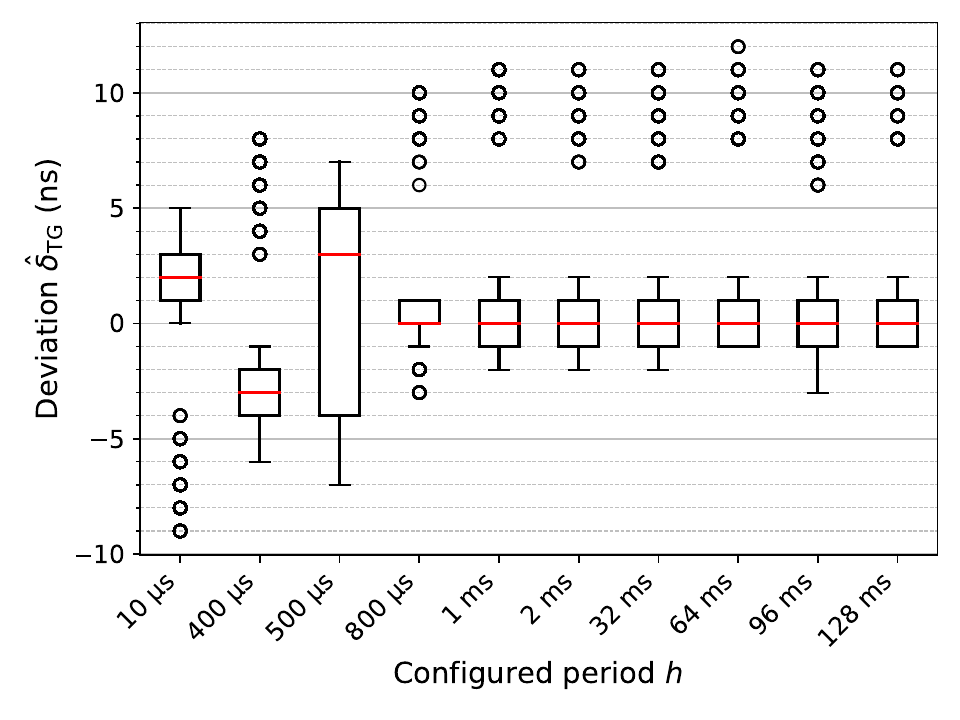}{Packet generator deviation $\hat{\delta}_{\text{TG}}$ for different configured periods.}

The boxplot in \fig{pdfs/boxplot_clock_offset} shows the median as a red line, the first and third quartiles as the edges of the box, and whiskers that extend to 1.5 times the interquartile range.
Values outside this range are plotted as outliers.
A positive $\hat{\delta}_{\text{TG}}$ indicates that the actual period exceeded the configured value by $\hat{\delta}_{\text{TG}}$ while a negative value means it was shorter by that amount.

Most periods show deviations within $\hat{\delta}_{\text{TG}} = \pm \qty{2}{\ns}$, with all outliers staying within $\pm$\qty{11}{\ns}.
An exception occurs at a period of \qty{500}{\us} which shows a wider spread up to $\pm \qty{5}{\ns}$ with fewer outliers.
We attribute this to internal scheduling behavior of the packet generator in the Intel Tofino\texttrademark\ switching ASIC.

Although these deviations are small, they can impact the periodicity computation.
If a period-completion frame arrives late, the computed relative position within the current \ac{GCL} cycle may exceed the period $h$, which would index an out-of-period entry.
To ensure that all frames are assigned to a valid \ac{tGCL} entry, P4-TAS clamps any calculated position $\ge h$ to the final entry of the cycle.
Conversely, if a period-completion frame arrives early, the periodicity mechanism in \sect{impl_periodicity} semantically evaluates the position modulo $h$, so the result always lies in $[0,h)$.
Therefore, the deviation from the configured period is compensated and all frames are mapped to existing \ac{GCL} entries.

\subsubsection{Queue Opening Delay}
\label{sec:eval_queue_delay}
In the \ac{TNA}, there is a small but non-zero delay between writing the \ac{AFC} value, i.e., between initiating a queue state change, and the actual update of the queue state in the hardware~\cite{SoKh23}.
To quantify internal delays in the \ac{AFC} mechanism, we measure the time between issuing a queue state change and the actual release of \ac{TSN} frames.
We denote this queue opening\footnote{Measurements showed that queue opening and closing delays are distributed in the same way in the TNA.} delay as $\hat{\delta}_{\text{queue}}$.
This delay impacts \ac{TAS} precision and is rarely documented in available hardware.
The measurement procedure is implemented in the data plane of P4-TAS and is shown in \fig{pdfs/queueing_delay}.

\figeps[0.9\columnwidth]{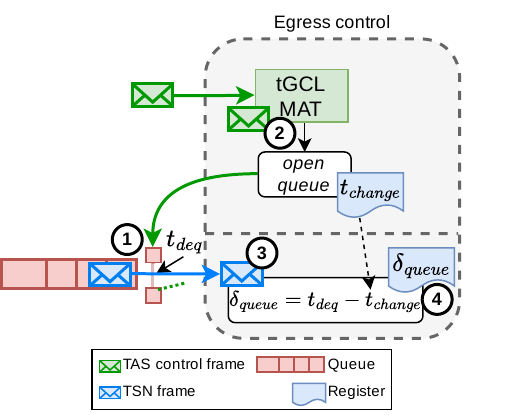}{Measurement of the queue opening delay in the data plane of P4-TAS.}

In \fig{pdfs/queueing_delay}, a closed queue is first filled with \ac{TSN} frames (step~\whiteballnumber{1}).
When a \ac{TAS} control frame matches a \ac{tGCL} entry that requires opening the queue, it triggers the opening via \ac{AFC} in the matched action and records the timestamp $t_\text{change}$ (step~\whiteballnumber{2}).
The dequeuing timestamp $t_\text{deq}$ of the first \ac{TSN} frame leaving the queue is then used to compute $\hat{\delta}_\text{queue}$ as shown in \equa{delta_queue} (step \whiteballnumber{3}):

\begin{align}
    \hat{\delta}_\text{queue} &= t_\text{deq} - t_\text{change}\label{eq:delta_queue}.
\end{align}

This value is stored as a time series in a register of the data plane for all observed transitions (step~\whiteballnumber{4}).

The \ac{tGCL} for this measurement is configured with eight consecutive entries, one per priority.
Each entry opens the corresponding priority queue for \qty{100}{\us}, so that the schedule cycles through all eight priorities in turn.
TSN traffic is generated using P4TG~\cite{p4tg,IhZi25Apr,IhZi25Sep} at \qty{400}{\gbps} with randomized priorities and \qty{64}{\byte} frames.
This ensures that the queues are saturated.
The experiment is run for \qty{60}{\second}.
\fig{pdfs/ccdf_queue_opening_delay} shows the \ac{CCDF} of the measured queue opening delay $\hat{\delta}_\text{queue}$.

\figeps[\columnwidth]{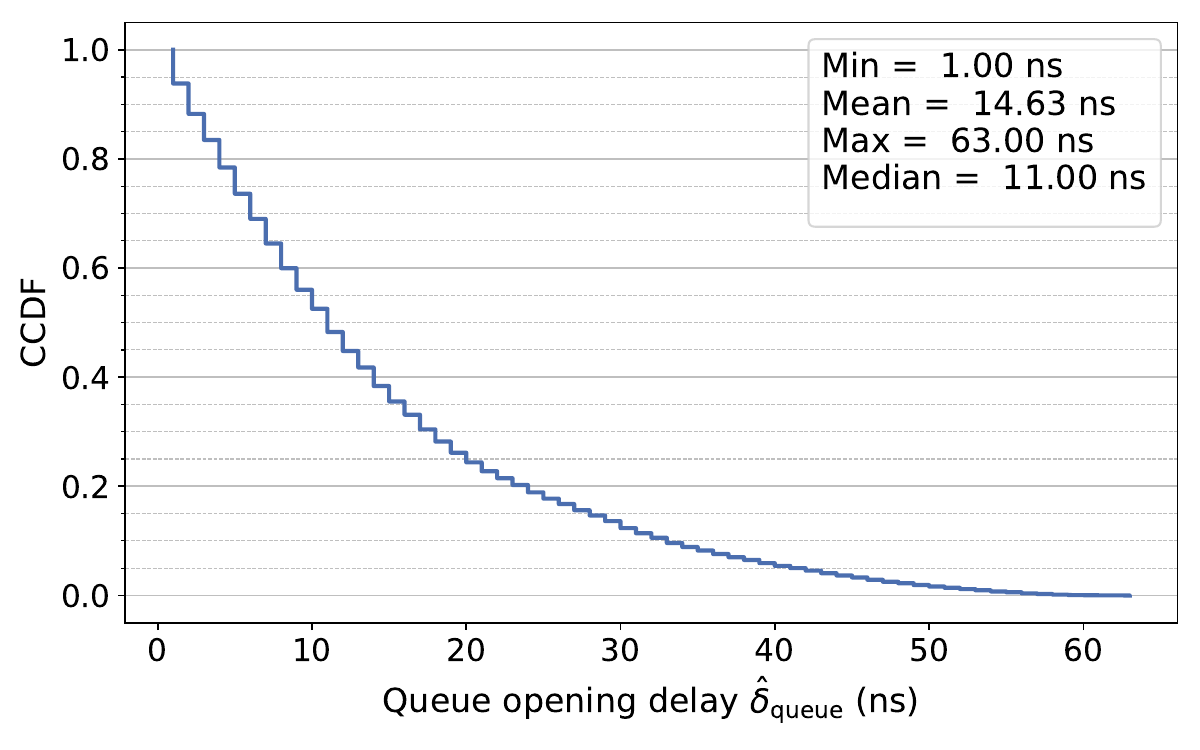}{\ac{CCDF} of measured queue opening delay.}

Most delays are below $\hat{\delta}_\text{queue} = \qty{11}{\ns}$ with a tail extending up to \qty{63}{\ns} and a mean of $\mu(\hat{\delta}_\text{queue}) = \qty{14.63}{\ns}$.
These results reveal small but measurable internal delays.
In particular, the queue opening delay can cause transitional behavior at \ac{tGCL} boundaries where frames from the previous entry may still be transmitted briefly after the next entry has started.
The impact of this effect and the role of \acfp{GSI} are evaluated in \sect{guard_bands}.

\subsubsection{TAS Control Traffic Delay}
\label{sec:eval_batch_delays}
For \ac{TAS} control frames, the internal packet generator is configured to generate a frame every nanosecond.
The frames are sequentially generated in batches of eight, with each frame controlling one of the eight priority queues.
In practice, however, a frame cannot be generated every nanosecond.
Instead, a small delay occurs between frame generation which limits the granularity at which queue state updates can be triggered.
To quantify this phenomenon, we collect the timestamp of each \ac{TAS} control frame in the data plane of P4-TAS and compute the delay $\hat{\delta}_{\text{control}}$ between two consecutive frames $i$ and $i+1$:

\begin{align}
    \hat{\delta}_{\text{control}} &= t_{i+1} - t_i.\label{eq:delta_intra}
\end{align}

We collect 100000 values for $\hat{\delta}_{\text{control}}$, all calculated in the data plane and stored in a time series register.
The resulting histogram is shown in \fig{pdfs/histogram_control-delay}.

\figeps[\columnwidth]{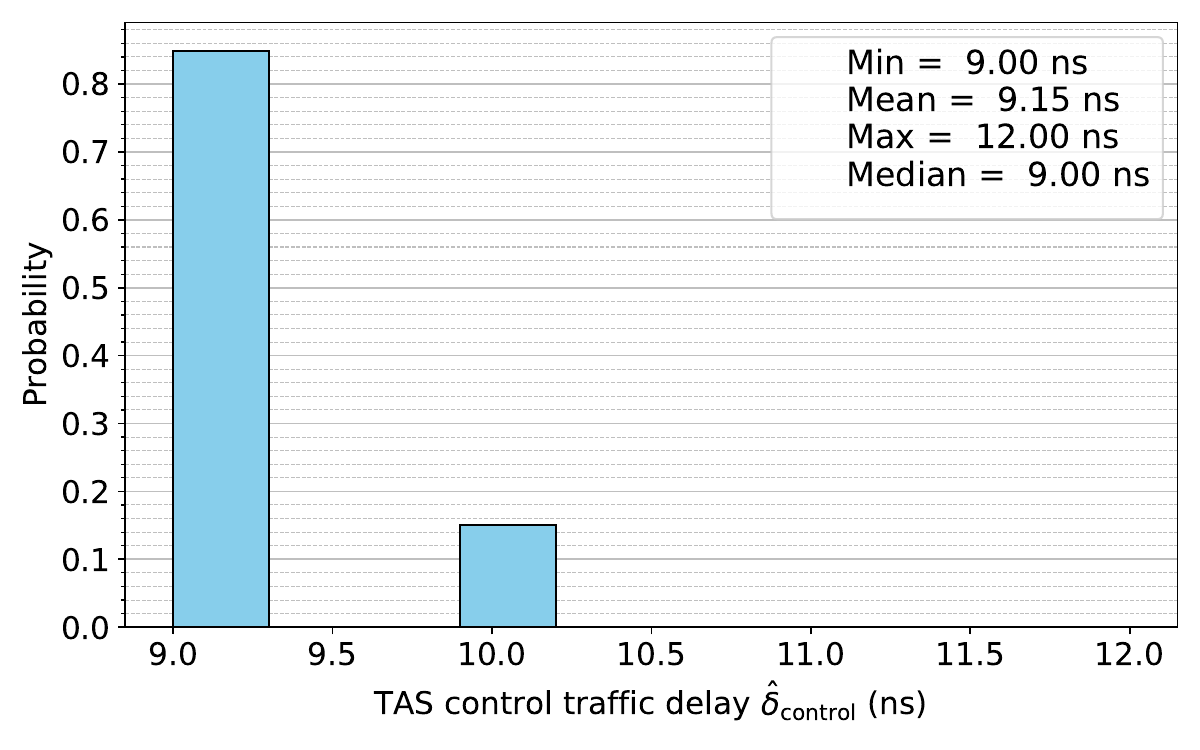}{Measured \ac{TAS} control traffic delay.}

The measured median is $\hat{\delta}_{\text{control,M}} = \qty{9}{\ns}$, with only a few frames showing a slightly higher delay of up to \qty{12}{\ns}.
Thus, queue states, i.e., transmission gates, can be updated only every \qty{9}{\ns}.
Because frames are generated sequentially in batches of eight, updates for different priority queues are offset sequentially by \qty{9}{\ns} and cannot occur simultaneously.
Further, this means that the transmission gate state update of the same priority can be triggered every $8 \cdot \hat{\delta}_{\text{control}} \approx \qty{72}{\ns}$.
This value should be seen as a worst-case upper bound.
In practice, the effective delay can be close to zero if a control frame arrives just before a scheduled gate change.
Such a short delay only matters if the \ac{tGCL} entry resolution is on the order of \qty{72}{\ns} which is much smaller than typical \ac{tGCL} entry durations~\cite{StOs24}.

\subsubsection{Summary of Internal Delays}
\label{sec:eval_summary_delays}
\tabl{internal_delays} gives an overview of the identified and measured internal delays in the best and in the worst case.

\begin{table}[htb!]
    \caption{Measured internal delays in P4-TAS.}
    \label{tab:internal_delays}
    \begin{tabularx}{\columnwidth}{|X|X|X|}
        \hline
        \textbf{Delay} & \textbf{Best case} &  \textbf{Worst case} \\
        \hline \hline
        $\hat{\delta}_{\text{TG}}$     & \qty{0}{\ns} & \qty{-11}{\ns} / \qty{11}{\ns} \\
        \hline
        $\hat{\delta}_{\text{queue}}$  & \qty{1}{\ns} & \qty{63}{\ns} \\
        \hline
        %$\delta_{\text{inter}}$  & \qty{0}{\ns} & \qty{83}{\ns} \\
        %\hline
        %$\delta_{\text{intra}}$  & \qty{0}{\ns} & \qty{12}{\ns} \\
        %\hline
        $\hat{\delta}_{\text{control}}$  & \qty{0}{\ns} & \qty{12}{\ns} \\
        \hline        
    \end{tabularx}
\end{table}

Those internal delays accumulate to $\Delta_{\text{internal}}$ shown in \equa{Delta_internal}:

\begin{align}
    \Delta_{\text{internal}} &= \delta_{\text{TG}} + \delta_{\text{queue}} + \delta_{\text{control}}\label{eq:Delta_internal}.
\end{align}

The internal delay $\Delta_{\text{internal}}$ may reduce or extend the actual duration of a \ac{tGCL} entry compared to its configured duration.
\fig{pdfs/delta_internal} illustrates this effect for three consecutive \ac{tGCL} entries of configured duration $d$.

\figeps[\columnwidth]{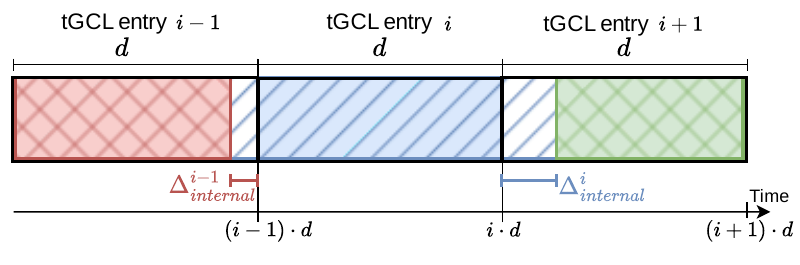}{Worst case effect of the internal delay $\Delta_{\text{internal}}$ on the \ac{tGCL} entry duration.}

If the preceding \ac{tGCL} entry $i-1$ experiences a negative internal delay, it is shortened while \ac{tGCL} entry $i$ is extended.
In addition, \ac{tGCL} entry $i$ itself may experience a positive delay.
In this case, the actual duration of \ac{tGCL} entry $i$ becomes

\begin{align}
    \hat{d}_i = d + |\Delta^{i-1}_{\text{internal}}| + \Delta^{i}_{\text{internal}}\label{eq:Delta_internal_cor}.
\end{align}

In the worst case, $\Delta^{i}_{\text{internal}}$ is composed of the maximum traffic generator deviation, queue opening delay, and control traffic delay: $\Delta^{i}_{\text{internal},\max} = \qty{11}{\ns} + \qty{63}{\ns} +\qty{12}{\ns}= \qty{86}{\ns}$.
Further, $\Delta^{i-1}_{\text{internal}}$ can be negative if the traffic generator deviation is negative and all other delays are close to zero, yielding up to \qty{11}{\ns} of shortening.
This implies that a \ac{tGCL} entry may be extended in duration by up to \qty{86}{\ns}, or be shortened by \qty{11}{\ns}.
Further, through correlation of consecutive \ac{tGCL} entries, a \ac{tGCL} entry may be extended by up to \qty{97}{\ns} as shown in \fig{pdfs/delta_internal}.

In the best case, a \acs{TAS} control traffic frame arrives exactly at the switchover point to a new \ac{tGCL} entry, resulting in a control traffic delay of $\delta_{\text{control},\min} = \qty{0}{\ns}$.
Combined with the measured best case queue delay $\delta_{\text{queue},\min} = \qty{1}{\ns}$, and traffic generator accuracy $\delta_{\text{TG,min}} = 0$, the best case internal delay is $\Delta_{\text{internal},\min} = \qty{1}{\ns}$.

These bounds can be incorporated into scheduling algorithms that account for hardware-induced uncertainty.
For example, Stüber \textit{et al.}~\cite{StOs24} propose a robust scheduling approach that models per-node processing delays as bounded intervals to account for hardware jitter, such as gate-opening delays.
The internal delays measured in this work provide exactly such bounds, while also existing in other hardware~\cite{EpOs25,sparx_1}.
However, this work validates mechanism-level timing behavior rather than network-level schedule synthesis.
Designing and evaluating concrete schedules that incorporate these values is outside the scope of this work.

\subsection{External Measurement of tGCL Entry Duration}
\label{sec:eval_accuracy}
Hardware-based \ac{TAS} implementations may show deviations between configured and actual \ac{tGCL} entry durations due to the internal delays characterized in \sect{eval_internal_delays}.
Such deviations have been observed in commercial TSN switches~\cite{EpOs25} but are rarely documented by vendors.
In this section, we describe a methodology to externally measure the duration of \ac{tGCL} entries.
First, we define the requirements for the methodology, present the testbed, and describe the measurement procedure.
Then, we apply it to P4-TAS, analyze the results, and introduce \acp{GSI} to improve timing accuracy.

\subsubsection{Methodology Requirements}
\label{sec:eval_requirements}
The measurement methodology determines the actual duration of tGCL entries compared to their configured values.
For that purpose, it detects time slice boundaries through priority transitions in the shaped output traffic on a dedicated measurement switch.
We apply the methodology to P4-TAS to verify consistency with the internal delay bounds from \sect{eval_internal_delays}.
Since the methodology relies solely on observing output traffic, it can be applied to other TAS implementations, including commercial black-box devices, without requiring access to internal state.
The methodology requires three conditions: (i) a \ac{tGCL} in which each time slice opens exactly one priority queue, (ii) saturated traffic to ensure precise boundary detection, with priorities randomly sampled from a uniform distribution, and (iii) a measurement device with hardware timestamping.
The one-priority-per-entry configuration is specifically designed for measurement purposes, as it allows time slice boundaries to be detected through priority changes in the output stream.
However, the underlying queue state change mechanism is identical regardless of how many queues are open simultaneously.
Therefore, the measured deviations equally apply to other \ac{tGCL} configurations.
The traffic rate determines how accurately boundaries can be measured.
A higher packet rate increases the measurement precision, as time slice boundaries can be resolved more closely.

\subsubsection{Testbed}
\label{sec:eval_testbed}
The testbed for the external \ac{tGCL} entry measurement is shown in \fig{pdfs/evaluation}.

\figeps[\columnwidth]{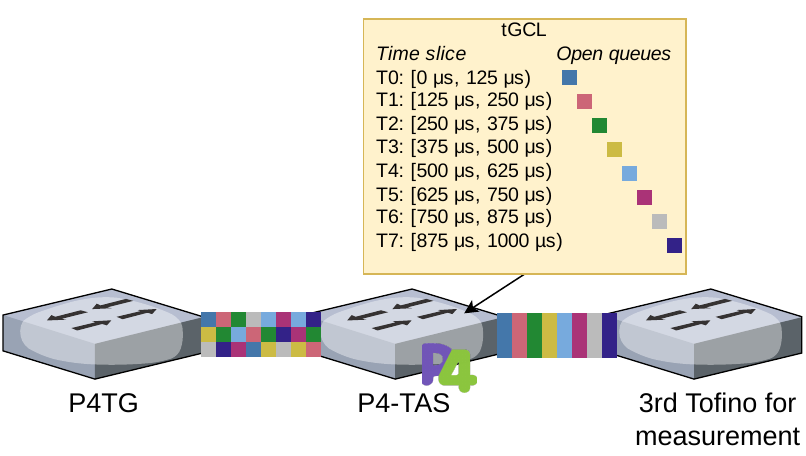}{Testbed for the external \ac{tGCL} entry measurement.}

A \ac{tGCL} with a period of \qty{1}{\ms} divided into eight \qty{125}{\us} entries is configured in P4-TAS.
During each entry, only one of the eight queues is open, corresponding to one priority. 
Traffic is generated with P4TG~\cite{p4tg,IhZi25Apr,IhZi25Sep} at a rate of \qty{557}{Mpps} using minimum-size \qty{64}{\byte} frames.
The high packet rate of \qty{557}{Mpps} results in a frame inter-arrival time of approximately \qty{2}{\ns}.
This provides precise boundary detection for the measurement procedure described in the following section.
P4TG assigns a random priority sampled from a uniform distribution to each frame and encapsulates it with MPLS to additionally test the DetNet translation.
P4-TAS translates incoming MPLS traffic into a TSN stream, after which the configured tGCL is applied based on the resulting \ac{TSN} stream identifier.
After shaping by the \ac{TAS}, the traffic is forwarded to a third Tofino\texttrademark\ switch which records frame arrival times per priority in a dedicated P4 program.

\subsubsection{Measurement Procedure}
\label{sec:eval_measurement_procedure}
The measurement procedure in the dedicated P4 program on the third switch is based on detecting changes in priority within the received stream.
The traffic generator sends a stream with randomized frame priorities.
This traffic is then shaped by the configured \ac{tGCL} in P4-TAS such that only frames of one priority are forwarded during each tGCL entry.
The measurement assumes that frames of only one priority $\pi \in \{0,\ldots, 7\}$ arrive at the measurement switch during each \ac{tGCL} entry.
A series of timestamps of the first and last frame in a \ac{tGCL} entry, i.e., of the same priority, is collected and stored in the data plane of the measurement switch.
The measurement is illustrated in \fig{pdfs/eval_measurement}.

\figeps[\columnwidth]{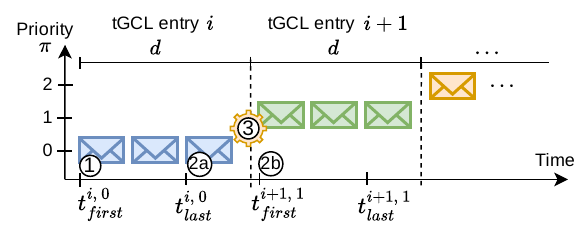}{Measurement procedure in the dedicated P4 program of the third Tofino switch.}

For priority $\pi=0$, the arrival time of the first frame in the $i$-th \ac{tGCL} entry is stored as $t^{i,\pi=0}_{\text{first}}$ in step \whiteballnumber{1}.
When the next priority $\pi=1$ appears, the arrival time of the last frame of the previous priority $\pi=0$ is stored as $t^{i,\pi=0}_{\text{last}}$ (step \whiteballnumber{2a}), and the new frame marks $t^{i+1,\pi=1}_{\text{first}}$ (step \whiteballnumber{2b}).
The control plane calculates the duration of entry $i$ for priority $\pi$ in step \whiteballnumber{3} as follows:

\begin{align}
    \hat{d}_i^\pi = t^{i,\pi}_{\text{last}} - t^{i,\pi}_{\text{first}}.
\end{align}

The measured \ac{tGCL} entry duration is then compared with the configured \ac{tGCL} entry duration of $d$ (\qty{125}{\us} in our configuration), and the deviation $\hat{\delta}^{i,\pi}_{\text{slice}}$ is obtained as

\begin{align}
    \hat{\delta}^{i,\pi}_{\text{slice}} =\hat{d}_i^\pi - d\label{eq:delta_slice}.
\end{align}

The same procedure is applied for all priorities.
A negative value for $\hat{\delta}^{i,\pi}_{\text{slice}}$ thus means that the measured \ac{tGCL} entry duration was shorter than the configured duration while a positive value means that it was longer.
A total of 32764 values for $\hat{\delta}^{i,\pi}_{\text{slice}}$ is collected per run.
The experiment is repeated ten times.

\subsubsection{Introducing Gate Switching Intervals (GSIs)}
\label{sec:guard_bands}
The internal queue opening/closing delay identified in \sect{eval_queue_delay} causes queue state transitions to occur during a short interval instead of instantaneously.
This may cause transitional behavior where queues of a \ac{tGCL} entry are not yet closed while queues of the next \ac{tGCL} entry have already begun forwarding.
As a result, frames from consecutive \ac{tGCL} entries may be transmitted simultaneously, violating the configured \ac{tGCL}.
This overlap is a phenomenon of the TAS, not an artifact of the measurement, and must be addressed.
To mitigate this effect, we introduce \acp{GSI} which are illustrated in \fig{pdfs/gsi}.

\figeps[\columnwidth]{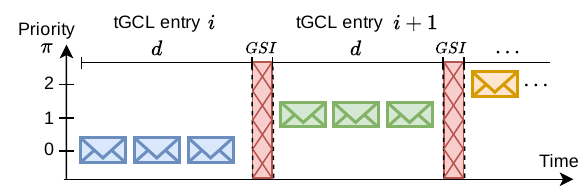}{\Acp{GSI} mitigate transitional behavior of \ac{tGCL} entries by explicitly closing all queues for a short time.}

\acp{GSI} are short, explicit \ac{tGCL} entries in which all queues are closed.
They are inserted between \ac{tGCL} entries.
These \acp{GSI} suppress transitional forwarding behavior and isolate each \ac{tGCL} entry.
We added \acp{GSI} of \qty{30}{\ns} which was sufficient to eliminate overlap without significantly impacting available transmission time.
While the worst-case queue opening delay measured in \sect{eval_internal_delays} reaches \qty{63}{\ns}, a \qty{30}{\ns} \ac{GSI} provides sufficient isolation because the \ac{GSI} itself is subject to the same internal delays.
This effectively extends the \ac{GSI}'s duration and ensures that queue state transitions complete before the next scheduled entry begins.
Larger \acp{GSI} did not improve the results.

Since \acp{GSI} are additional \ac{tGCL} entries, they add to the overall \ac{tGCL} duration.
Two options exist for accommodating the \acp{GSI}.
First, the period can be extended by the cumulative \ac{GSI} duration to preserve the configured time slice durations.
In this case, a \ac{tGCL} with $n$ entries and a configured period $h$ results in an effective period of $h+n\cdot d_\text{GSI}$.
Second, if the application requires a fixed period, the \ac{GSI} duration can be subtracted from individual time slices, shortening each by $d_\text{GSI}$.
In that case, the available transmission time per time slice is reduced, which may cause frames to be delayed to the subsequent period.
The choice depends on whether the application constrains the period or the per-slice transmission budget.
In our evaluation, \acp{GSI} extend the period.
Accordingly, the \qty{1}{\ms} period configured in \sect{eval_testbed} becomes \qty{1.00024}{\ms} with \acp{GSI} of \qty{30}{\ns}.
 
\subsubsection{Measurement Results}
First, we measured the deviation of the observed values from the configured duration of \ac{tGCL} entries without introducing \acp{GSI}.
The resulting statistics were unreliable, with a mean deviation across all measurements of $\mu(\hat{\delta}_{\text{slice}}) = \qty{-58592}{\ns}$ and a median of \qty{-999}{\ns}.
These apparent deviations are not meaningful because consecutive entries frequently overlapped at their boundaries.
This violates the measurement assumption that only frames of a single priority arrive during each \ac{tGCL} entry.
After introducing \acp{GSI}, the mean deviation improves to $\mu(\hat{\delta}_{\text{slice}})= \qty{-18}{\ns}$ with a median of \qty{-16}{\ns}.

The distribution of deviations with \acp{GSI}, averaged across all ten runs, is presented in \fig{pdfs/equal-intervals-1ms_one-gcl_30ns-gb_64-B}.
The error bars indicate 95\% confidence intervals across the ten runs, showing low variance between repetitions.
The metric is computed per priority, i.e., for each $\pi$ and entry $i$ as $\hat{\delta}^{i,\pi}_{\text{slice}}$.
The histogram aggregates all priorities because the behavior is identical across them.

\figeps[\columnwidth]{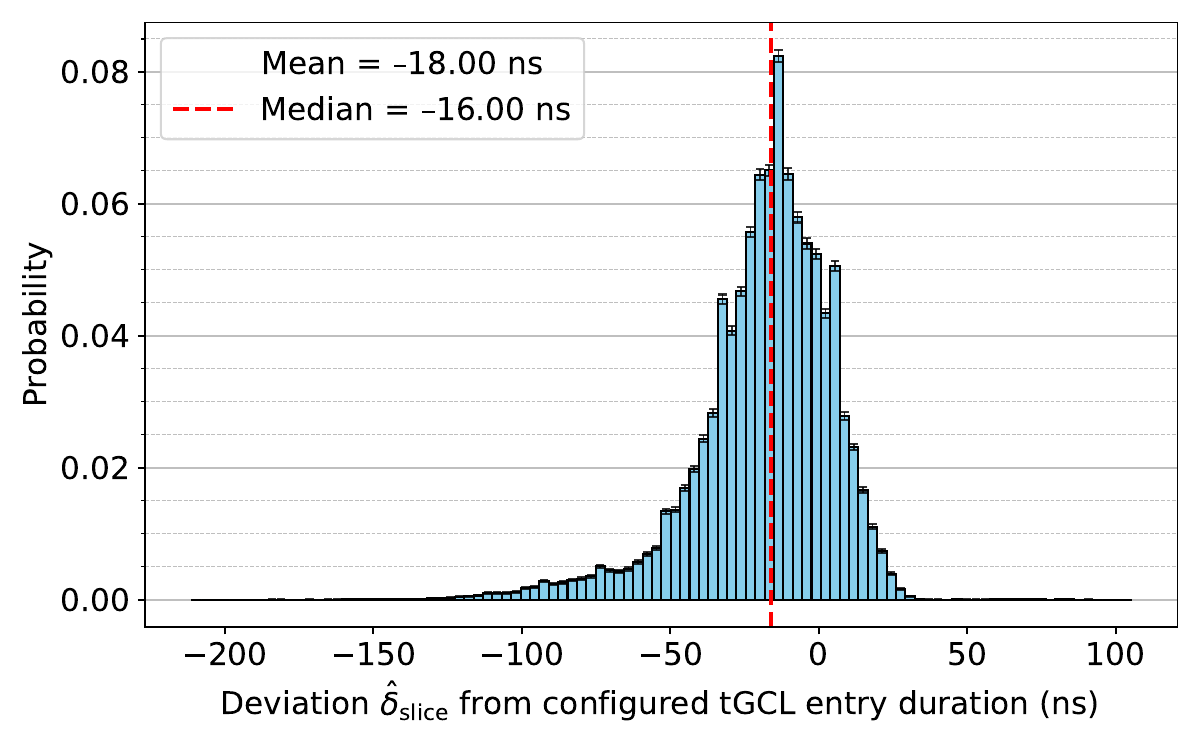}{Normalized histogram of the deviation between configured and measured \ac{tGCL} entry duration with \acp{GSI}.}

A negative deviation means the measured \ac{tGCL} entry duration was shorter than configured, while a positive deviation means it was longer. 
Positive deviations up to approximately \qty{50}{\ns} occur when internal delays extend a \ac{tGCL} entry beyond its configured duration.
Negative deviations result from the correlated effect of consecutive boundary delays: if a preceding \ac{tGCL} entry is extended by a positive delay, the following \ac{tGCL} entry starts late and is consequently shortened.
The distribution is centered around a mean of \qty{-18}{\ns}.
This systematic negative bias is attributed to the \acp{GSI}.
Since \acp{GSI} are explicit \ac{tGCL} entries, they are subject to the same internal delays as regular \ac{tGCL} entries.
In particular, the queue opening delay causes the \ac{GSI} to remain closed slightly longer than configured, which shortens the subsequent \ac{tGCL} entry.
The observed bias of \qty{-18}{\ns} is consistent with the mean queue opening delay of \qty{14.63}{\ns} measured in \sect{eval_queue_delay}.
Despite this bias, \acp{GSI} are essential to isolate consecutive \ac{tGCL} entries and prevent the boundary overlap that renders measurements without \acp{GSI} unreliable.
The observed deviations are consistent with the internal delay bounds from \sect{eval_internal_delays}.
Positive deviations stay within the \qty{86}{\ns} worst-case internal delay, while negative deviations result from the combined effect of \ac{GSI} shortening and correlation with preceding \ac{tGCL} entries.

We repeated the measurement for periods of \qty{400}{\us}, \qty{800}{\us}, and \qty{4}{\ms}.
Across all configurations, median deviations range from \qty{-12}{\ns} to \qty{-25}{\ns}.
This confirms that the deviations remain in the same range regardless of the configured period.
Further, these results demonstrate that actual \ac{tGCL} entry durations deviate from their configured values in practice and confirm the results by Eppler \textit{et al.}~\cite{EpOs25} and Stüber \textit{et al.}~\cite{StOs24}.
This reinforces the need for scheduling algorithms to account for hardware-induced timing deviations~\cite{StOs24}.

\subsubsection{Cost of Gate Switching Intervals}
\label{sec:eval_gsi}
While \acp{GSI} are essential to isolate consecutive \ac{tGCL} entries, they have two associated costs: reduced effective transmission time and additional \ac{MAT} entries.

With \qty{30}{\ns} \acp{GSI} and eight time slices of \qty{125}{\us} as configured in \sect{eval_testbed}, each \ac{GSI} consumes 0.024\% of a time slice.
For shorter time slices, the relative overhead increases but remains small: at \qty{50}{\us} per time slice, it is 0.06\%, and even at \qty{1.25}{\us} (a \qty{10}{\us} period with eight entries), it reaches only 2.4\%.
The overhead exceeds 5\% only for time slices shorter than approximately \qty{600}{\ns}, which are already impractical given the \qty{86}{\ns} worst case internal delay.
In absolute terms, \qty{30}{\ns} of lost transmission time corresponds to about one minimum-size (\qty{64}{\byte}) frame at \qty{25}{\gbps} and approximately 18 frames at \qty{400}{\gbps}.

\acp{GSI} also add additional \ac{tGCL} entries to the MAT, with the exact number depending on the range-to-ternary conversion overhead.
For the configuration evaluated in \sect{eval_testbed} ($n=8, h=\qty{1}{\ms}$), \acp{GSI} require 416 additional ternary MAT entries.
This overhead consumes approximately 1\% of the total \ac{tGCL} \ac{MAT} capacity.

Therefore, \acp{GSI} impose a minimal cost while providing the essential benefit of cleanly isolating consecutive time slices.

\subsection{Scalability}
\label{sec:scalability}
Scalability is a critical aspect for \ac{TSN} and \ac{DetNet} deployments which often involve large numbers of scheduled traffic streams.
However, many scheduling algorithms overlook hardware resource constraints such as limited \ac{MAT} capacity~\cite{StOs24}.
In this section, we evaluate the scalability of our P4-TAS implementation by analyzing the number of supported \ac{tGCL} and \ac{sGCL} entries, and the number of streams for DetNet and TSN stream identification.

\subsubsection{Range-to-Ternary Conversion Overhead}
\label{sec:eval_r-to-t}
In this section, we evaluate the required number of \ac{MAT} entries for \acp{tGCL} and \acp{sGCL}.
We first explain how many \ac{MAT} entries are required to model a \ac{GCL}.
Then, we derive a worst-case upper bound for the required number of ternary entries using the range-to-ternary conversion.
Finally, we measure the actual number of ternary \ac{MAT} entries for representative \ac{tGCL} and \ac{sGCL} configurations.

\paragraph{Required Number of MAT Entries per GCL Entry}
\acp{GCL} are periodically repeated lists of entries with each entry consisting of a time slice and gate states.
Since one \ac{tGCL} entry encodes eight queue states, a \ac{tGCL} with $n$ entries requires
\begin{align}
    n_{\text{range}}^{\text{tGCL}} &= 8\cdot n
\end{align}
 range \ac{MAT} entries, each matching on one of the queues and the time slice range.
 
An \ac{sGCL} entry only encodes one stream gate state.
Therefore, an \ac{sGCL} with $n$ entries requires
\begin{align}
     n_{\text{range}}^{\text{sGCL}} &= n
\end{align} 
range \ac{MAT} entries, i.e., one per time slice.
P4-TAS employs a range-to-ternary conversion described in \sect{range-to-ternary} to enable matching on the entire timestamp.
The conversion algorithm replaces a single range \ac{MAT} entry with multiple ternary \ac{MAT} entries, which increases the number of required MAT entries.
The \ac{tGCL} \ac{MAT} supports 39000 ternary entries while the \ac{sGCL} \ac{MAT} holds 6000 ternary entries.
\acp{sGCL} are configured per stream gate and may be shared by multiple streams.

\paragraph{Worst Case Upper Bound}
We derive a worst case upper bound for the number of required ternary \ac{MAT} entries in \acp{GCL} through the range-to-ternary conversion.
Let $w$ denote the bit width of the timestamp field used as the match key in the \ac{MAT}. 
Although the timestamp field has \qty{48}{bits} in our implementation, all slice boundaries lie within a period, i.e., within $[0,h)$.
Thus, the effective width is bounded by $w = \lceil \log_2(h) \rceil$, since $h \ll 2^{48}$.
Gupta \textit{et al.}~\cite{GuMc01} showed that a range of width $w$ bits can be transformed into at most $n_{\text{ternary,max}} = 2 \cdot w - 2$ ternary entries.
Consequently, in the worst case, modeling a \ac{tGCL} and a \ac{sGCL} containing $n$ \ac{GCL} entries results in the following:
\begin{align}
    n_{\text{ternary,max}}^{\text{tGCL}} &= 8 \cdot n \cdot (2\cdot w - 2)\label{eq:tgcl}\\[6pt]
    n_{\text{ternary,max}}^{\text{sGCL}} &= n \cdot (2\cdot w - 2)\label{eq:sgcl}.
\end{align}
In practice, the actual number is often significantly lower due to favorable alignment.
For example, if the range's width is a power of two and properly aligned, a single ternary entry may suffice.

\paragraph{Measured Ternary Entry Counts}
We evaluate the number of ternary \ac{MAT} entries required after range-to-ternary conversion for representative \ac{sGCL} and \ac{tGCL} configurations.
We vary the period $h \in \{\qty{10}{\us}, \qty{400}{\us}, \qty{500}{\us}, \qty{524}{\us}\,(2^{19}\,\mathrm{ns}), \qty{2}{\ms}, \qty{128}{\ms}\}$ and the number of \ac{GCL} entries $n \in \{2,3,4,5,8,10,16,20,32\}$.
For each configuration, the period is divided into $n$ equally sized time slices.
The results are shown in \fig{pdfs/tas_entries_vs_slices_no_gb}.

\figeps[\columnwidth]{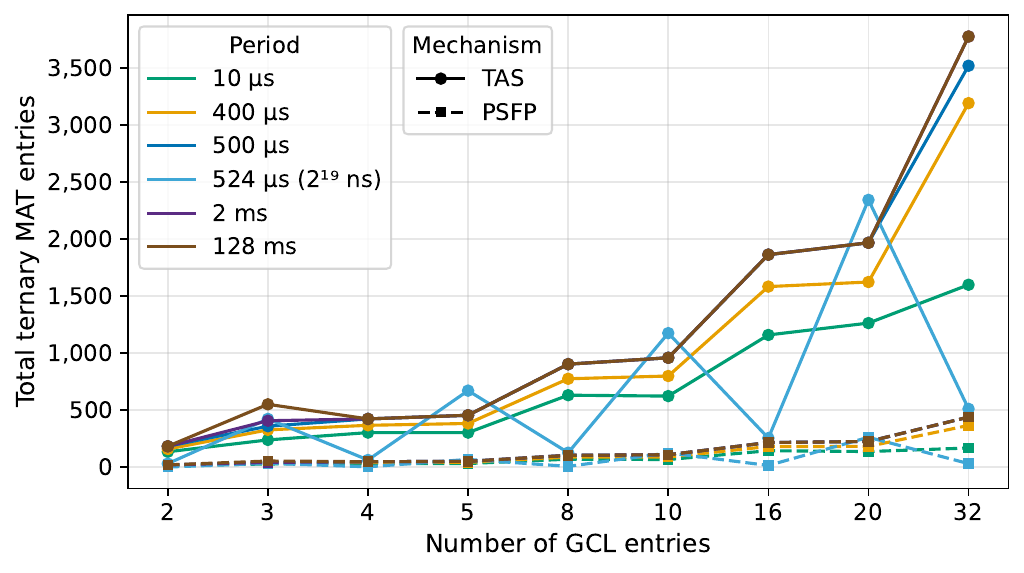}{Number of required ternary entries for different \ac{sGCL} and \ac{tGCL} configurations.}

Generally, the required number of ternary entries in \fig{pdfs/tas_entries_vs_slices_no_gb} increases with the period and the number of \ac{GCL} entries.
Across the evaluated configurations, the total number of ternary entries ranges from 136 (TAS, $h=\qty{10}{\us}$, $n=2$) up to 3776 (TAS, $h=\qty{128}{\ms}$, $n=32$), and from 15 (PSFP, $h=\qty{10}{\us}$, $n=2$) up to 440 (PSFP, $h=\qty{128}{\ms}$, $n=32$).

\input{chapters/comparison_table}

Power-of-two aligned configurations substantially reduce the required number of entries.
For example, a \ac{TAS} configuration with $h=\qty{500}{\us}$ and $n=32$ requires 3520 ternary entries, while the power-of-two aligned period $h=\qty{524}{\us}$ ($2^{19}$ ns) with $n=32$ requires only 512 ternary entries.
This behavior is expected because the conversion produces fewer ternary prefixes when slice boundaries align with power-of-two blocks.
Given the 39000 available \ac{tGCL} \ac{MAT} entries, even the largest evaluated configuration with 3776 entries uses less than 10\% of this capacity, leaving room for multiple concurrent \acp{tGCL}.
However, if \ac{MAT} capacity becomes a bottleneck, a power-of-two-aligned period and slice boundaries can be configured.
In addition, future work can apply optimized range-to-ternary conversion algorithms to further reduce ternary entry counts~\cite{SuKi10, BrHa18, SuKi10_2, DoSu6}.

\subsubsection{Number of DetNet and TSN Streams}
To evaluate the scalability in terms of streams in our implementation, we analyze the structure and capacity of the \ac{MAT} used for \ac{DetNet} and \ac{TSN} stream identification.

A single MAT handles identification of \ac{DetNet} and \ac{TSN} streams.
It uses ternary keys consisting of the S-Label for \ac{DetNet} streams, and Ethernet destination address, VLAN ID and IPv4 source and destination address for \ac{TSN} streams~\cite{cb}.
The use of ternary match entries enables wildcarding and aggregation.
For example, an entry matching only on the S-Label enables DetNet-to-TSN translation while another matching on MAC destination and VLAN ID supports TSN-to-DetNet translation and \ac{TSN} stream identification.

The \ac{MAT} supports 8196 entries which allows more than 8196 \ac{DetNet} or \ac{TSN} streams to be identified.
In cases where IP-based identification is used, ternary aggregation can further increase the number of identifiable streams.
A survey by Stüber \textit{et al.}~\cite{StOs23} reports deployments with up to 10812 streams, indicating that our implementation can support realistic industrial-scale scenarios with appropriate use of wildcarding.
These results show that the design is scalable and capable of supporting a number of streams typical in TSN/DetNet deployments.

\subsection{Comparison of TAS-Capable Platforms}
\label{sec:eval_compare}

In this section, we summarize and compare capabilities of several TAS-capable platforms, including our P4-TAS prototype on a P4-programmable ASIC.
The overview is shown in \tabl{tas_hw_comparison}.

Similar to P4-TAS, the Predict6G open-source platform provides \ac{TAS} and \ac{DetNet} integration, but its documentation does not specify configurable time resolution, internal delay behavior, or scalability~\cite{MeOl25}.
Commercial platforms such as NXP's SJA1105TEL~\cite{nxp_1, nxp_2}, Microchip's SparX-5i family~\cite{sparx_1} and PD-IES008~\cite{microchip_pd1,microchip_pd2}, and Relyum's RELY-TSN12~\cite{relyum} provide hardware support for \ac{TAS} and \ac{PSFP}.
However, publicly available specifications stop at time granularity, queue counts, or \ac{GCL} sizes while omitting internal delay sources that ultimately determine schedule precision.
Although these devices advertise nanosecond-level configuration granularity, our evaluation with P4-TAS shows that practical gate updates are constrained by internal delays in the range of tens of nanoseconds.
This phenomenon is further supported by Eppler \textit{et al.}~\cite{EpOs25} who report internal TAS delays of approximately \qty{2.6}{\micro\second} in Relyum switches.
This demonstrates that internal \ac{TAS} timing effects are inherent to the mechanism itself and not specific to P4-based implementations, and can be significantly larger in proprietary devices.
Since such delays are rarely documented, the effective precision of commercial solutions is difficult to assess from datasheets alone.
Further, those delays are internal and cannot usually be isolated directly in commercial black-box switches.

Our P4-TAS prototype achieves a comparable time configuration granularity of \qty{1}{\ns} while also documenting measured internal delays.
Specifically, we observed a worst-case internal delay for a \ac{tGCL} entry of $\Delta_{\text{internal,max}} = \qty{86}{\ns}$ in the evaluation.
In contrast, vendor platforms either do not document such values (e.g., NXP SJA1105TEL, Microchip PD-IES008) or only disclose partial information (e.g., SparX-5i which specifies a queue opening delay of $\delta_{\text{queue}} = \qty{512}{\ns}$).
The transparency in P4-TAS allows a more realistic assessment of achievable schedule precision.
In terms of scalability, P4-TAS supports a larger number of flows ($\geq$8196) and larger \acp{GCL} (39k for TAS and 6k for PSFP) compared to the commercial platforms.
For \ac{GCL} entries, the range-to-ternary conversion overhead must be considered which we evaluated in \sect{eval_r-to-t}.

A further distinction is line rate throughput.
While most commercial TSN-capable switch ASICs target 1–25~Gb/s per port in automotive and industrial domains, P4-TAS operates at up to 400~Gb/s per port.
This enables its use not only in TSN deployments but also in high-speed data center environments where integration with \ac{DetNet} becomes relevant.
Hence, P4-TAS extends the design space beyond today’s embedded and industrial use cases.

Overall, the comparison shows that commodity hardware already supports \ac{TAS} and \ac{PSFP} functionality, but vendors disclose little about their internal timing behavior.
This lack of transparency makes it difficult to design schedules with nanosecond accuracy.
P4-TAS fills this gap by explicitly characterizing internal delays, enabling more predictable and transparent use of \ac{TSN}.

%% file: chapters/comparison_table.tex
\begin{table*}[htb!]
  \caption{Comparison of TAS-capable platforms.}
  \label{tab:tas_hw_comparison}
  \setlength{\tabcolsep}{3pt}
  \renewcommand{\arraystretch}{1.05}
  \small
  \begin{tabularx}{\textwidth}{|Y|Y|c|c|c|c|c|Y|Y|Y|Y|}
    \hline
    \textbf{Model} & \textbf{Type} & \textbf{Line rate} &
    \multicolumn{3}{c|}{\textbf{Features}} &
    \makecell{\textbf{Configurable}\\\textbf{time resolution}} &
    \makecell{\textbf{Internal}\\\textbf{delays}} &
    \makecell{\textbf{Number of}\\\textbf{streams}} &
    \multicolumn{2}{c|}{\textbf{GCL size}} \\ \cline{4-6}\cline{10-11}
     & & & \textbf{TAS} & \textbf{PSFP} & \makecell{\textbf{DetNet}\\\textbf{integration}} & & & & \textbf{TAS} & \textbf{PSFP} \\ \hline\hline

    P4-TAS (Tofino 2) & Hardware (prog. ASIC) & 32$\times$400 Gb/s &
    \yes & \yes & \yes & 1 ns &
    \makecell[l]{$\Delta_{\text{internal, max}}=$\\ 86 ns} &
    $\ge$ 8196 & 39k entries & 6k entries \\ \hline

    \makecell[l]{Predict6G\\Open Source\\TSN Platform}\cite{MeOl25} & Software (XDP) & $\le$ 1 Gb/s &
    \yes & \no & \yes & n/a & n/a & n/a & n/a & n/a \\ \hline    

    NXP SJA1105TEL\cite{nxp_1,nxp_2} & Hardware (ASIC) & 5$\times$1 Gb/s &
    \yes & \yes & \no & 200 ns & n/a & 1024 & n/a & n/a \\ \hline

    Microchip SparX-5i family~\cite{sparx_1} & Hardware (ASIC) & 8$\times$25 Gb/s &
    \yes & \yes & \no & 1 ns &
    $\delta_{\text{queue}}=$512 ns~\cite{sparx_1} &
    $\ge$ 256 & \makecell[l]{10k entries} &
    1024 stream gates with 4 entries each \\ \hline

    Microchip PD-IES008~\cite{microchip_pd1,microchip_pd2} & Hardware (n/a) & 8$\times$1 Gb/s &
    \yes & \yes & \no & 1 ns & n/a & n/a & 8 tGCLs with 256 entries each & n/a \\ \hline

    Relyum RELY-TSN12~\cite{relyum} & Hardware (FPGA) & 8$\times$1 Gb/s &
    \yes & \yes & \no & n/a & $\delta_{\text{queue}}\approx \qty{2.6}{\us}$~\cite{EpOs25}
     & n/a & n/a & n/a \\ \hline

  \end{tabularx}
\end{table*}

%% file: chapters/07-conclusion.tex
\section{Conclusion}
\label{sec:conclusion}
We presented P4-TAS, a P4-based implementation of the \ac{TAS} for \ac{TSN} on the Intel Tofino\texttrademark\ 2.
P4-TAS introduces a novel mechanism for periodic queue state control.
For that purpose, it uses a continuous stream of internally generated \ac{TAS} control frames to trigger queue state updates according to the \ac{tGCL}.
To the best of our knowledge, this enables TAS functionality on a P4-programmable ASIC for the first time.
P4-TAS further provides an MPLS/TSN translation layer that bridges TSN and DetNet domains.
This extends traffic shaping and policing to routed Layer 3 networks at line rates up to \qty{400}{\gbps}.
Additionally, P4-TAS incorporates \ac{PSFP} as well as IEEE~Std~1588 \ac{PTP} synchronization with ordinary clock and transparent clock support.
We improved our earlier P4-PSFP design by eliminating recirculation and increasing the \ac{GCL} time resolution to the nanosecond scale using a range-to-ternary algorithm.

We identified and quantified three sources of delay in the TAS mechanism: traffic generator inaccuracy, queue opening delay, and control frame delay.
Our evaluations show a worst-case accumulated delay of about \qty{86}{\ns} between \ac{tGCL} entries.
This worst-case value is substantially smaller than the microsecond-scale gate transition delays reported for some commercial TSN switches~\cite{EpOs25,sparx_1}.
These delays affect the precision of gate transitions rather than user traffic.
While the specific values are platform-dependent, the underlying delay sources, particularly the queue opening delay, are inherent to hardware-based TAS implementations in general~\cite{EpOs25}.

We proposed a measurement methodology to externally measure \ac{TAS} time slice accuracy.
The methodology captures priority transitions in the shaped output traffic on a dedicated measurement switch to determine the actual duration of individual time slices.
The results confirmed consistency between internally measured and externally observed deviations.
They further revealed overlap of consecutive time slices caused by non-instantaneous queue state changes.
This phenomenon is independently confirmed by Eppler \textit{et al.}~\cite{EpOs25} on commercial TSN switches.
As a mitigation, we introduced \acp{GSI}, short explicit \ac{tGCL} entries in which all queues are closed.

The \ac{PTP} evaluation confirmed synchronization accuracy within tens of nanoseconds across multiple hops, and the scalability analysis demonstrates support for 39000 ternary \ac{tGCL} \ac{MAT} entries and more than 8196 flows, covering the requirements of current industrial deployments.

Compared with existing ASIC- and FPGA-based TSN platforms, P4-TAS offers similar configurability in terms of time granularity, but additionally exposes internal delays that directly affect scheduling precision.
This transparency allows schedules to be designed with awareness of hardware-induced deviations, difficult to achieve with today’s black-box hardware.
Moreover, P4-TAS supports line rates up to \qty{400}{\gbps} per port and seamless \ac{DetNet}/\ac{TSN} translation, extending applicability from industrial and automotive networks to high-throughput environments such as data centers and carrier backbones.

Future work will focus on improving scalability, for example by optimizing range-to-ternary usage, and on investigating how delay characterization can be incorporated into scheduling algorithms to increase robustness against hardware-level variability.
Further, we will explore the integration of P4-TAS into a \ac{PTP}-synchronized multi-hop \ac{TSN} testbed to validate gate scheduling and latency guarantees under realistic \ac{TSN}-specific traffic patterns.

%% file: glossar.tex
\section*{List of Abbreviations}
\begin{acronym}[F-Label]
    \acro{AFC}{advanced flow control}
    \acro{ATS}{asynchronous traffic shaper}
    \acro{CBS}{credit-based shaper}
    \acro{CCDF}{complementary cumulative distribution function}
    \acro{d-CW}{DetNet control word}
    \acro{DetNet}{deterministic networking}
    \acro{E2E}{end-to-end}
    \acro{F-Label}{forward label}
    \acro{GSI}{gate switching interval}
    \acro{GCL}{gate control list}
    \acro{MAT}{match+action table}
    \acro{OC}{ordinary clock}
    \acro{P2P}{peer-to-peer}
    \acro{P4}{programming protocol-independent packet processors}
    \acro{PSFP}{per-stream filtering and policing}
    \acro{PTP}{precision time protocol}
    \acro{QoS}{quality of service}
    \acro{RSVP}{resource reservation protocol}
    \acro{sGCL}{stream GCL}
    \acro{S-Label}{service label}
    \acro{TAS}{time-aware shaper}
    \acro{TDMA}{time division multiple access}
    \acro{tGCL}{transmission GCL}
    \acro{TNA}{tofino native architecture}
    \acro{TC}{transparent clock}
    \acro{TSN}{time-sensitive networking}
\end{acronym}

%% file: conferences.bib
@STRING{ SIGCOMM = {{ACM SIGCOMM}}}

@STRING{ Networks = {{International Telecommunication Network Strategy and Planning Symposium (Networks)}}}

@STRING{ Networking = {{IFIP-TC6 Networking Conference (Networking)}}}

@STRING{ CoNEXT = {{ACM Conference on emerging Networking EXperiments and Technologies (CoNEXT)}}}

@STRING{ JNCA = {{Journal of Network and Computer Applications (JNCA)}}}

@STRING{ CCNC = {{IEEE Consumer Communications and Networking Conference (CCNC)}}}

@STRING{ ICTON = {{International Conference on Transparent Optical Networks (ICTON)}}}

@STRING{CNSM = {{International Conference on Network and Services Management (CNSM)}}}

@STRING{ Performance = {International Symposium on Computer Performance, Modeling, Measurements and Evaluation (Performance)}}

@STRING{ ETFA = {{IEEE International Conference on Emerging Technologies and Factory Automation (ETFA)}}}

@STRING{ WFCS = {{IEEE International Workshop on Factory Communication Systems (WFCS)}}}

@STRING{ NetSoft = {{IEEE Conference on Network Softwarization (NetSoft)}}}

@STRING{ ToN   = {IEEE\slash ACM Transactions on Networking}}

@STRING{ Computer   = {{IEEE} Computer Magazine}}

@STRING{ CCR = {{ACM SIGCOMM Computer Communication Review}}}

@STRING{ TNSM = {{{IEEE} Transactions on Network and Service Management}}}

@misc{ns,
   title     = {{UCB/LBNL/VINT Network Simulator -- ns (Version 2)}},
   howpublished = {Source code and documentation available at
       http://www-mash.cs.berkeley.edu/ns/}
}


%% file: literature.bib
@ARTICLE{cb,
	journal={IEEE Std 802.1CB}, 
	title={{IEEE Standard for Local and Metropolitan Area Networks--Frame Replication and Elimination for Reliability}}, 
	year={2017},
	volume={},
	number={}}

@ARTICLE{qci,
  author={},
  journal={IEEE Std 802.1Qci}, 
  title={{IEEE Standard for Local and Metropolitan Area Networks--Bridges and Bridged Networks--Amendment 28: Per-Stream Filtering and Policing}}, 
  year={2017},
  volume={},
  number={}}

@ARTICLE{qcc,
  author={},
  journal={IEEE Std 802.1Qcc}, 
  title={{IEEE Standard for Local and Metropolitan Area Networks--Bridges and Bridged Networks -- Amendment 31: Stream Reservation Protocol (SRP) Enhancements and Performance Improvements}}, 
  year={2018},
  volume={},
  number={}}

@ARTICLE{as,
  author={},
  journal={IEEE Std 802.1AS}, 
  title={{IEEE Standard for Local and Metropolitan Area Networks--Timing and Synchronization for Time-Sensitive Applications}}, 
    year=2020,
  volume={},
  number={}}

@ARTICLE{8021q,
  author={},
  journal={IEEE Std 802.1Q}, 
  title={{IEEE Standard for Local and Metropolitan Area Networks -- Bridges and Bridged Networks}}, 
  volume={},
  number={},
  year=2022}

@ARTICLE{qbv,
  journal={IEEE Std 802.1Qbv}, 
  title={{IEEE Standard for Local and Metropolitan Area Networks -- Bridges and Bridged Networks - Amendment 25: Enhancements for Scheduled Traffic}}, 
  year={2016},
  volume={},
  number={}}

@ARTICLE{8021cbdb,
  author={},
  journal={{IEEE Std 802.1CBdb}}, 
  title={{IEEE Standard for Local and Metropolitan Area Networks--Frame Replication and Elimination for Reliability Amendment 2: Extended Stream Identification Functions}}, 
  year={2022},
    month = mar}

@ARTICLE{qav,
  author={},
  journal={IEEE Std 802.1Qav}, 
  title={{IEEE Standard for Local and Metropolitan Area Networks-- Virtual Bridged Local Area Networks Amendment 12: Forwarding and Queuing Enhancements for Time-Sensitive Streams}}, 
  year=2019,
  volume={},
  number={}}

@article{kn,
	title = {{A Survey on Data Plane Programming with P4: Fundamentals, Advances, and Applied Research}},
	journal = JNCA,
	volume = 212,
	year = 2023,
    month = mar,
	author = {Frederik Hauser and Marco Häberle and Daniel Merling and Steffen Lindner and Vladimir Gurevich and Florian Zeiger and Reinhard Frank and Michael Menth},
}

@misc{p4spec,
	author = {{The P4 Language Consortium}},
	title = {{{$P4_{16}$} Language Specification}},
	month = may,
	year = 2021,
	note = "\textit{Last accessed on 14.05.2023}",
	howpublished = {\url{https://p4.org/p4-spec/docs/P4-16-v1.2.2.pdf}},
}

@article{p4,
    author = {Bosshart, Pat and Daly, Dan and Gibb, Glen and Izzard, Martin and McKeown, Nick and Rexford, Jennifer and Schlesinger, Cole and Talayco, Dan and Vahdat, Amin and Varghese, George and Walker, David},
    title = {{P4: Programming Protocol-Independent Packet Processors}},
    year = 2014,
    volume = 44,
    number = 3,
    journal = CCR,
    month = jul,
    numpages = 9,
}

@misc{tna,
	author = {{Intel{$^{\circledR}$}}},
	title = {{{$P4_{16}$} Intel{$^{\circledR}$} Tofino™ Native Architecture – Public Version}},
	month = apr,
	year = 2021,
	note = "\textit{Last accessed on 07.06.2023}",
	howpublished = {\url{https://github.com/barefootnetworks/Open-Tofino/blob/master/PUBLIC_Tofino-Native-Arch.pdf}},
}

@ARTICLE{p4tg,
author={Lindner, Steffen and Häberle, Marco and Menth, Michael},
journal={{IEEE Access}}, 
title={{P4TG: 1 Tb/s Traffic Generation for Ethernet/IP Networks}}, 
year=2023,
month = feb,
volume=11,
pages={17525--17535}}

@ARTICLE{StOs23,
  author={Stüber, Thomas and Osswald, Lukas and Lindner, Steffen and Menth, Michael},
  journal={IEEE Access}, 
  title={{A Survey of Scheduling Algorithms for the Time-Aware Shaper in Time-Sensitive Networking (TSN)}}, 
  year=2023,
  month = jun,
  volume=11,
  pages={61192--61233}
}

@INPROCEEDINGS{CrOl21,
  author={Craciunas, Silviu S. and Oliver, Ramon Serna},
  booktitle=WFCS, 
  title={{Out-of-Sync Schedule Robustness for Time-Sensitive Networks}}, 
  year=2021,
  mon = jun,
  pages={75--82}}

@ARTICLE{NaTh19,
  author={Nasrallah, Ahmed and Thyagaturu, Akhilesh S. and Alharbi, Ziyad and Wang, Cuixiang and Shao, Xing and Reisslein, Martin and ElBakoury, Hesham},
  journal={IEEE Communications Surveys \& Tutorials}, 
  title={{Ultra-Low Latency (ULL) Networks: The IEEE TSN and IETF DetNet Standards and Related 5G ULL Research}}, 
  year=2019,
  month = sep,
  volume=21,
  number=1,
  pages={88--145}}

@misc{git,
	author = {{Chair of Communication Networks, University of Tuebingen}},
	title = {{GitHub: P4-TAS}},
	howpublished = {\url{https://github.com/uni-tue-kn/P4-TAS}},
}

@misc{rfc9037,
    series =    {Request for Comments},
    number =    9037,
    author =    {Balazs Varga and János Farkas and Andrew G. Malis and Stewart Bryant},
    title =     {{Deterministic Networking (DetNet) Data Plane: MPLS over IEEE 802.1 Time-Sensitive Networking (TSN)}},
    pagetotal = 11,
    year =      2021,
    month =     jun,
}

@article{AbGh23,
  title={{Integration of DetNet/TSN Reliability Functions in 5G Systems: A Case Study and Measurements}},
  author={Mohammed A. Abuibaid and Amir Hoseein Ghorab and Aysun Aslan Saruhan and Marc St-Hilaire and Glenn Parsons and J{\'a}nos Farkas and Bal{\'a}zs Varga and Istv{\'a}n Moldov{\'a}n and Mikl{\'o}s M{\'a}t{\'e} and Syed Hassan Raza Naqvi},
  journal={IEEE Conference on Standards for Communications and Networking (CSCN)},
  year={2023},
  pages={369--375},
}

@ARTICLE{IhLi24,
  author={Ihle, Fabian and Lindner, Steffen and Menth, Michael},
  journal=TNSM, 
  title={{P4-PSFP: P4-Based Per-Stream Filtering and Policing for Time-Sensitive Networking}}, 
  year=2024,
  volume={21},
  number={5},
  pages={5273--5290},
}

@misc{rfc8655,
    series =    {Request for Comments},
    number =    8655,
    howpublished =  {RFC 8655},
    publisher = {RFC Editor},
    author =    {Norman Finn and Pascal Thubert and Balazs Varga and János Farkas},
    title =     {{Deterministic Networking Architecture}},
    pagetotal = 38,
    year =      2019,
    month =     oct,
}

@misc{rfc8964,
    series =    {Request for Comments},
    number =    8964,
    publisher = {RFC Editor},
    doi =       {10.17487/RFC8964},
    author =    {Balazs Varga and János Farkas and Lou Berger and Andrew G. Malis and Stewart Bryant and Jouni Korhonen},
    title =     {{Deterministic Networking (DetNet) Data Plane: MPLS}},
    pagetotal = 27,
    year =      2021,
    month =     jan,
}

@misc{rfc9024,
    series =    {Request for Comments},
    number =    9024,
    howpublished =  {RFC 9024},
    publisher = {RFC Editor},
    author =    {Balazs Varga and János Farkas and Andrew G. Malis and Stewart Bryant and Don Fedyk},
    title =     {{Deterministic Networking (DetNet) Data Plane: IEEE 802.1 Time-Sensitive Networking over MPLS}},
    pagetotal = 12,
    year =      2021,
    month =     jun,
}

@misc{rfc8939,
    series =    {Request for Comments},
    number =    8939,
    howpublished =  {RFC 8939},
    publisher = {RFC Editor},
    doi =       {10.17487/RFC8939},
    author =    {Balazs Varga and János Farkas and Lou Berger and Don Fedyk and Stewart Bryant},
    title =     {{Deterministic Networking (DetNet) Data Plane: IP}},
    pagetotal = 21,
    year =      2020,
    month =     nov,
}

@INPROCEEDINGS{AdIa20,
  author={Addanki, Vamsi and Iannone, Luigi},
  booktitle={IFIP Networking Conference (Networking)}, 
  title={{Moving a Step Forward in the Quest for Deterministic Networks (DetNet)}}, 
  year={2020},
  volume={},
  number={},
  pages={458--466}}

@INPROCEEDINGS{HeGe16,
  author={Heise, Peter and Geyer, Fabien and Obermaisser, Roman},
  booktitle={IFIP International Conference on New Technologies, Mobility and Security (NTMS)}, 
  title={{TSimNet: An Industrial Time Sensitive Networking Simulation Framework Based on OMNeT++}}, 
  year={2016},
  volume={},
  number={}}

@INPROCEEDINGS{FaHe19,
  author={Falk, Jonathan and Hellmanns, David and Carabelli, Ben and Nayak, Naresh and Dürr, Frank and Kehrer, Stephan and Rothermel, Kurt},
  booktitle={International Conference on Networked Systems (NetSys)}, 
  title={{NeSTiNg: Simulating IEEE Time-sensitive Networking (TSN) in OMNeT++}}, 
  year={2019},
  volume={},
  number={}}

@INPROCEEDINGS{JiLi19,
  author={Jiang, Junhui and Li, Yuting and Hong, Seung Ho and Xu, Aidong and Wang, Kai},
  booktitle={IEEE International Conference on Mechatronics and Automation (ICMA)}, 
  title={{A Time-Sensitive Networking (TSN) Simulation Model Based on OMNET++}}, 
  year={2018},
  pages={643--648}}

@ARTICLE{AhAk24,
  author={Ben Ahmed, Akram and Hirofuchi, Takahiro and Fukai, Takaaki},
  journal={IEEE Access}, 
  title={{Hardware Design and Evaluation of an FPGA-Based Network Switch Supporting Asynchronous Traffic Shaping for Time Sensitive Networking}}, 
  year={2024},
  volume={12},
  number={},
  pages={123149--123165}}

@INPROCEEDINGS{AkHi24,
  author={Ben Ahmed, Akram and Hirofuchi, Takahiro and Fukai, Takaaki},
  booktitle={IEEE International Conference on Emerging Technologies and Factory Automation (ETFA)}, 
  title={{FPGA-Based Network Switch Architecture Supporting Credit Based Shaper for Time Sensitive Networks}}, 
  year={2024},
  volume={},
  number={}}

@article{BeHe22,
  title={{Analyzing and Modeling the Latency and Jitter Behavior of Mixed Industrial TSN and DetNet Networks}},
  author={Lukas W{\"u}steney and David Hellmanns and Markus Schramm and Lukas Osswald and Ren{\'e} Hummen and Michael Menth and Tobias Heer},
  journal=CoNext,
  year={2022},
}

@article{BrHa18,
  author={Bremler-Barr, Anat and Harchol, Yotam and Hay, David and Hel-Or, Yacov},
title = {{Encoding Short Ranges in TCAM Without Expansion: Efficient Algorithm and Applications}},
year = {2018},
volume = {26},
number = {2},
journal = {IEEE ToN},
month = apr,
pages = {835-–850},
}

@inproceedings{SuKi10,
author = {Sun, Yan and Kim, Min Sik},
title = {{Bidirectional Range Extension for TCAM-based Packet Classification}},
year = {2010},
booktitle = {IFIP-TC6 Networking},
pages = {351-–361},
}

@INPROCEEDINGS{SuKi10_2,
  author={Sun, Yan and Kim, Min Sik},
  booktitle={IEEE CCNC}, 
  title={{Tree-Based Minimization of TCAM Entries for Packet Classification}}, 
  year={2010},
  volume={},
  number={},
  pages={1--5},
}

@inproceedings{DoSu6,
author = {Dong, Qunfeng and Banerjee, Suman and Wang, Jia and Agrawal, Dheeraj and Shukla, Ashutosh},
title = {{Packet Classifiers in Ternary CAMs Can Be Smaller}},
year = {2006},
booktitle = {ACM SIGMETRICS/IFIP PERFORMANCE},
pages = {311-–322},
}

@ARTICLE{GuMc01,
  author={Gupta, P. and McKeown, N.},
  journal={IEEE Network}, 
  title={{Algorithms for Packet Classification}}, 
  year={2001},
  volume={15},
  number={2},
  pages={24--32},
}

@phdthesis{Su11,
author = {Sun, Yan},
advisor = {Kim, Min Sik},
title = {{Scalable Packet Processing for High-speed Networks}},
year = {2011},
school = {Washington State University},
publisher = {Washington State University},
}

@ARTICLE{PoCi25,
  author={Polverini, Marco and Cianfrani, Antonio and Caiazzi, Tommaso and Scazzariello, Mariano},
  journal={IEEE Journal on Selected Areas in Communications}, 
  title={{SRv6 Meets DetNet: A New Behavior for Low Latency and High Reliability}}, 
  year={2025},
  volume={43},
  number={2},
  pages={448--458}}

@ARTICLE{StEp24,
  author={Stüber, Thomas and Eppler, Manuel and Osswald, Lukas and Menth, Michael},
  journal={IEEE Transactions on Industrial Informatics}, 
  title={{Performance Comparison of Offline Scheduling Algorithms for the Time-Aware Shaper (TAS)}}, 
  year={2024},
  volume={20},
  number={7},
  pages={9736--9748}}

@article{FrZa24,
title = {{A Comprehensive Latency Profiling Study of the Tofino P4 Programmable ASIC-based Hardware}},
journal = {Computer Communications},
volume = {218},
pages = {14--30},
year = {2024},
author = {David Franco and Eder {Ollora Zaballa} and Mingyuan Zang and Asier Atutxa and Jorge Sasiain and Aleksander Pruski and Elisa Rojas and Marivi Higuero and Eduardo Jacob},
}

@ARTICLE{StOs24,
  author={Stüber, Thomas and Osswald, Lukas and Menth, Michael},
  journal={IEEE Open Journal of the Communications Society}, 
  title={{Efficient Robust Schedules (ERS) for Time-Aware Shaping in Time-Sensitive Networking}}, 
  year={2024},
  volume={5},
  number={},
  pages={6655--6673}}

@inproceedings{DuYa24,
author = {Duan, Shihui and Yan, Renhe and Chen, Jie and Liu, Meihui and Xu, Qichen},
title = {{Comparison and Testing of Time Synchronization Accuracy between IEEE 1588v2 and IEEE 802.1AS}},
year = {2024},
booktitle = {International Conference on Communication and Network Security (ICCNS)},
pages = {300–-305},
}

@inproceedings{IhZi25Apr,
  author    = {Ihle, Fabian and Zink, Etienne and Lindner, Steffen and Menth, Michael},
  title     = {{Enhancements to P4TG: Protocols, Performance, and Automation}},
  year      = 2025,
  month     = apr,
  booktitle = {{KuVS Workshop on Network Softwarization (KuVS NetSoft)}}
}

@inproceedings{IhZi25Sep,
  author    = {Ihle, Fabian and Zink, Etienne and Menth, Michael},
  title     = {{Enhancements to P4TG: Histogram-Based RTT Monitoring in the Data Plane}},
  year      = 2025,
  month     = sep,
  booktitle = {{Workshop on Resilient Networks and Systems (ReNeSys)}}
}

@ARTICLE{ptp,
  author={},
  journal={IEEE Std 1588-2019}, 
  title={{IEEE Standard for a Precision Clock Synchronization Protocol for Networked Measurement and Control Systems}}, 
  year={2020}}

@inproceedings{SoKh23,
author = {Song, Cha Hwan and Khooi, Xin Zhe and Joshi, Raj and Choi, Inho and Li, Jialin and Chan, Mun Choon},
title = {{Network Load Balancing with In-network Reordering Support for RDMA}},
year = {2023},
booktitle = {ACM SIGCOMM},
pages = {816–-831},
numpages = {16},
}

@misc{nxp_1,
  author       = {{NXP Semiconductors}},
  title        = {{SJA1105 Product Data Sheet}},
  year         = {2016},
  month        = {Nov},
  url          = {\url{https://www.nxp.com/docs/en/data-sheet/SJA1105.pdf}},
  note = "\textit{Last accessed on 20.08.2025}",
}

@misc{nxp_2,
  author       = {{NXP Semiconductors}},
  title        = {{User Manual for Industry Linux Solution Release-v0.1}},
  year         = {2017},
  month        = {Jan},
  url          = {\url{https://www.nxp.com.cn/docs/en/reference-manual/IS-TSN-UM.pdf}},
  note = "\textit{Last accessed on 20.08.2025}",
}

@misc{sparx_1,
  author       = {{Microchip}},
  title        = {{SparX-5i Family of L2/L3 Industrial 25G Ethernet Switches}},
  year         = {2023},
  month        = {Nov},
  url          = {\url{https://ww1.microchip.com/downloads/aemDocuments/documents/UNG/ProductDocuments/DataSheets/SparX-5i%2B_Family_L2L3_Industrial_25G_Ethernet_Switches_Datasheet_DS00003757.pdf}},
  note = "\textit{Last accessed on 20.08.2025}",
}

@misc{microchip_pd1,
  author       = {{Microchip}},
  title        = {{PD-IES008/PD-IES008P/PD-IES206/PD-IES206P}},
  year         = {2025},
  month        = {Jan},
  url          = {\url{https://ww1.microchip.com/downloads/aemDocuments/documents/POE/ProductDocuments/Brochures/PD-IESxxx-family-v1-DS00004843.pdf}},
  note = "\textit{Last accessed on 20.08.2025}",
}

@misc{microchip_pd2,
  author       = {{Microchip}},
  title        = {{PD-IESxxx Manual}},
  year         = {2025},
  month        = {Aug},
  url          = {\url{https://onlinedocs.microchip.com/oxy/GUID-26305006-1311-4F15-9F7A-A91D9D7E2667-en-US-1/GUID-08AD29AB-F030-4C7C-8FA1-77102ABDA14E.html}},
  note = "\textit{Last accessed on 20.08.2025}",
}

@misc{relyum,
  author       = {{Relyum Industrial}},
  title        = {{Brochure: RELY-TSN12}},
  year         = {2025},
  month        = {Aug},
  url          = {\url{https://soc-e.com/wp-content/uploads/2024/09/BR_RELY-TSN-PCIe-w.pdf}},
  note = "\textit{Last accessed on 07.11.2025}",
}

@INPROCEEDINGS{MeOl25,
  author={Menendez, David Rico and De La Oliva, Antonio and Barroso-Fernández, Carlos and Schempp, Francisco Luque},
  booktitle={International Conference on Transparent Optical Networks (ICTON)}, 
  title={{Bridging the Gap between TSN and Open-Source}}, 
  year={2025},
  month=jul
}

@INPROCEEDINGS{EpOs25,
  author={Eppler, Manuel and Osswald, Lukas and Mitsakis, Nikolaos and Blenk, Andreas and Menth, Michael},
  booktitle=CoNext, 
  title={{T(SN)-Ray: Gauging TAS and PSFP Delays of TSN Switches for Predictable Deterministic Networking}}, 
  year={2025},
  month=dec
}

@ARTICLE{ieee1588,
  author={},
  journal={IEEE Std 1588}, 
  title={{IEEE Standard for a Precision Clock Synchronization Protocol for Networked Measurement and Control Systems}}, 
  year={2020},
}

@INPROCEEDINGS{ReHe22,
  author={Rezabek, Filip and Helm, Max and Leonhardt, Tizian and Carle, Georg},
  booktitle={International Conference on Network and Service Management (CNSM)}, 
  title={{PTP Security Measures and their Impact on Synchronization Accuracy}}, 
  year={2022},
  pages={109--117}}

@inproceedings{KySp19,
author = {Kyriakakis, Eleftherios and Spars\o{}, Jens and Schoeberl, Martin},
title = {{Hardware Assisted Clock Synchronization with the IEEE 1588-2008 Precision Time Protocol}},
year = {2018},
booktitle = {{International Conference on Real-Time Networks and Systems}},
pages = {51-–60}}
